\documentclass[a4paper,prb,reqno,twocolumn,showpacs,footinbib]{revtex4-2}
\usepackage[utf8]{inputenc}
\usepackage{amsmath}
\usepackage{amsfonts}
\usepackage{dsfont} 
\usepackage{amssymb}
\usepackage{makeidx}
\usepackage{graphicx}
\usepackage{color}
\usepackage{mathptmx} 
\usepackage{changes,enumitem,todonotes}
\usepackage{soul}
\newcommand{\bea}{\begin{eqnarray}}
\newcommand{\eea}{\end{eqnarray}}
\newcommand{\nn}{\nonumber}
\usepackage[mathscr]{euscript}
\usepackage{tikz}
\usepackage[normalem]{ulem}
\usepackage[utf8]{inputenc}
\usepackage{amsmath}
\usepackage{amsfonts}
\usepackage{amssymb}
\usepackage{subcaption} 
\usepackage{hyperref}
\hypersetup{colorlinks=true,linkcolor=blue,urlcolor=blue,citecolor=blue} 
\usepackage{mathtools} 
\usepackage{soul,xcolor}

\begin{document}
\setstcolor{red}
\title{Nonreciprocal electrical transport in linear systems with balanced gain and loss in the bulk}
\author{Rupak Bag and Dibyendu Roy}
\affiliation{ Raman Research Institute, Bengaluru 560080, India}



\begin{abstract}
 We investigate electrical transport in a quantum wire of $N$ sites connected to an equal number $(N_i/2)$ of sources and drains of charges in bulk. Each source and drain injects and extracts charges at the same rate, respectively. We show that the linear-response electrical current is nonreciprocal in such a system when the arrangement of sources and drains breaks the system's parity. We prove that inelastic scattering is essential for nonreciprocity in this system. For this, we invoke a master equation description of classical charge transport in a similar system. The nonreciprocal current in quantum wire matches that in the classical model for $N_i/N \sim 1$, generating a finite scattering length much smaller than the length of the wire. The nonreciprocity in the quantum wire oscillates with wire length when $N_i/N \ll 1$, and it can vanish at specific lengths.    
\end{abstract}
\maketitle

\section{Introduction}\label{Sec_Intro}
 Charge transport in matters has a long history, starting from empirical law by Ohm in the early days to kinetic equation description by Drude to recent fully quantum transport analysis pioneered by Landauer and B{\"u}ttiker \cite{Mello2004, Dutta2005, Akkermans2007}. Even today, electrical transport is one of the most important research topics for practical applications and basic understanding \cite{Badawy_RevElectronicTransportNanowire_ChemRev2024,  Lord_NanoContactNanowire_Nano2015,kumbhakar_UltraStrongElectronPhonon_arxiv2024,dutta_DiffusionLengthPhotoInducedChargeTransfer_2021, Fabian_FirstPrinciple_PhysRevB2022,Muller_CavityEnhancedChargeTranspot_PRL2017, Antipov_MajoranaNanonwires_PhysRevX2018,Pagano_IronSuperconductingNanowire_2020NanoMat, li2019probing, Islam_Pseudodiffusive_PhysRevResearch2020, Mirza_JunctionlessTransport_SciRep2017}. Here, we revisit an intriguing problem of charge transport in an open one-dimensional (1D) system, which is connected to the alternative sources (S's) and drains (D's) of charges in the middle. These sources and drains lead to the gain and loss of particles and energies in the system. The role of balanced gain and loss of particles or energies has been extensively investigated in recent years in the context of an effective parity-time $\mathcal{PT}$ symmetry in classical and quantum systems \cite{Bender_PT_PhysRevLett1998, Mostafazadeh_2002, Bender_2002_generalized, Ruter_PTinOptics_nphys2010,Schindler_PTinActiveLRC_PhysRevA2011,Bender_AsymmetricTransportActiveNonlinearity_PhysRevLett2013,Chang2014,Peng2014}. 
We examine the role of balanced injection and extraction of particles from a wire in classical and quantum transport regimes.  \textcolor{black}{Motivated by recent efforts to explore transport in effective $\mathcal{PT}$ symmetric Hamiltonian systems \cite{Ruter_PTinOptics_nphys2010,Schindler_PTinActiveLRC_PhysRevA2011,Chang2014,Peng2014}, we formulate a statistical description of transport in linear systems with balanced gain and loss of particles or charges in bulk. In contrast to most studies in effective $\mathcal{PT}$ symmetric Hamiltonian systems, our work does not require any untested assumptions in describing transport (e.g., use of a non-Hermitian effective Hamiltonian in the dynamical evolution applying the Schr{\"o}dinger or Heisenberg equations). Further, our statistical modeling of such an open quantum electrical system shows the explicit presence of quantum noises \cite{GSAgarwal_SpontaneousPhotons_PhysRevA2012, Roy2024}, which lacks in an effective closed system description. While directional transport has been demonstrated in effective $\mathcal{PT}$ symmetric models only in the nonlinear regime \cite{Chang2014,Peng2014}, our study shows the presence of balanced gain and loss can lead to nonreciprocal transport in the linear-response regime \cite{Roy2024}. This phenomenon can be explained within both quantum and classical descriptions, even without the magnetic fields.}

\begin{figure}[h!]
\includegraphics[height=1.7cm,width=8.4cm]{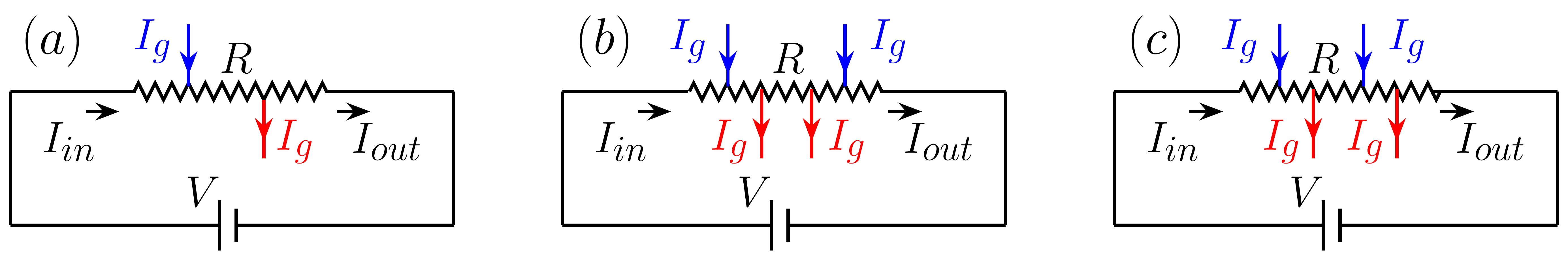}
\caption{\textcolor{black}{Three resistive circuits where the arrangement of current sources (blue arrow) and drains (red arrow) in (a) and (c) breaks the parity of the wire, but retains in (b).}}
\label{Fig0}
\end{figure}

 Let us consider a wire of length $L$ and resistance $R$, which is biased by voltage $V$ at the boundaries as shown in the inset of Fig.~\ref{Fig0}(a). The wire is further connected to a current S and a D at $L/3$ distance from the left and the right boundary, respectively. The S (D) injects (draws) an electric current $I_g$ in (from) the wire. The electric current flowing in $(I_{\text{in}})$ and out $(I_{\text{out}})$ at the boundaries of the wire can be calculated easily following the Ohm's or Kirchhoff's circuit law, and it is $I_{\text{in}}=I_{\text{out}}=(V/R-I_g/3)$ for a forward bias ($V>0$). The current values change to $(V/R+I_g/3)$ for a reverse bias ($V<0$).
 Thus, the charge transport through the wire in the presence of balanced gain and loss of charges from the S and D is nonreciprocal. The presence of S and D breaks the parity of the wire, which leads to the nonreciprocity in the current response. The role of breaking parity becomes more evident when we consider more than one pair of S and D within the circuit, e.g.,  we take two pairs of S and D connected to the wire as depicted in Figs.~\ref{Fig0}(b-c). The S-D configuration in Fig.~\ref{Fig0}(b) maintains the parity of the wire, but the parity is broken in Fig.~\ref{Fig0}(c). Here, the resistance in each segment between two consecutive connection points is $R/5$. Therefore, the electric current flowing into the circuit is $I_{\text{in}}=V/R$ in the former  (Fig.\ref{Fig0}(b)), and $I_{\text{in}}=(V/R-2I_g/5)$ in the latter (Fig.\ref{Fig0}(c)). Reversing the voltage bias across the circuits gives nonreciprocal transport only for the configuration in Fig.~\ref{Fig0}(c), where the parity is broken. Nevertheless, it is not clear what the physical mechanism/origin of the current nonreciprocity is. We seek to understand the mechanism in classical and quantum transport processes. We particularly reveal that the nonreciprocity in linear-response transport emerges for a finite rate of gain and loss in such systems in the presence of inelastic scattering and parity-breaking due to the sources and drains.   

\section{Master equation description}\label{Sec_MEq}

 We first invoke a master equation description (inspired by the Drude model) of charge $(e)$ transport with two species of right- and left-moving charges \cite{DRoy_ElectronSelfConsistent_PhysRevB2007}. Consider a 1D lattice model with $N$ sites and lattice spacing $a$. We take $N_i/2 \equiv (N/2-1)$ dimers of a pair of S and D connected to the bulk sites of the lattice. Each such S (D) is again injecting (extracting) $I_g$ charge current in (from) the system. We take $\rho^{\pm}(x)$ as the density of the right- and left-movers at position $x$ on the lattice. In bulk, these right- and left-movers hop with probability $p$ to the next site in the right and left direction, respectively. They can also convert between themselves with a finite probability $(1-p)$ at the connection points of the sources and drains with the lattice. In addition, the presence of S (D) current $I_g$ at position $x$ increases (reduces) the density of the left- and the right-movers by an amount of $I_g \tau/(2ea)$.  
\textcolor{black}{Such random motion of left- and right-movers with inter-conversion leads to the telegrapher’s equation in the absence of S's and D's \cite{DRoy_ElectronSelfConsistent_PhysRevB2007,PersistentRW_Goldstein1951,Kac1974,weiss_someTE_Physics_2002}.} To generate a bias across the system similar to the voltage difference, we connect the two ends of the lattice to two reservoirs. Thus, we set the density of particles inside the reservoirs as $\rho^{\pm}(x=0)=\rho_0+\delta \rho$ at the left terminal, and $\rho^{\pm}(x=(N+1)a)=\rho_0$ at the right terminal for a forward bias. For the reverse biasing, we switch the densities at two terminals. 
We then write discrete time-evolution equation for the density fields $\rho^{\pm}(x,t)$ in time steps $\tau$ as
\bea
&&\rho^{+}(a,t+\tau)-\rho^{+}(a,t)=\rho^{+}(0,t)-\rho^{+}(a,t), \label{me1}\\
&&\rho^{-}(a,t+\tau)-\rho^{-}(a,t)=p\rho^{-}(2a,t)-\rho^{-}(a,t), \label{me2}\\
&&\rho^{\pm}(x,t+\tau)-\rho^{\pm}(x,t)=p[\rho^{\pm}(x\mp a,t)-\rho^{\pm}(x,t)]\nn\\
&&~~~~-\frac{p-1}{2}[(1\pm 1)\delta_{x,2a}+(1\mp1)\delta_{x,(N-1)a}]\rho^{\pm}(x\mp a,t)\nn\\
&&~~~~+(1-p)[\rho^{\mp}(x,t)-\rho^{\pm}(x,t)]+(-1)^{\frac{x}{a}}\frac{I_g \tau}{2ea},\label{me3}\\
&&\rho^{+}(Na,t+\tau)-\rho^{+}(Na,t)=p\rho^{+}(L,t)-\rho^{+}(Na,t),\label{me4}\\
&&\rho^{-}(Na,t+\tau)-\rho^{-}(Na,t)=\rho^{-}(L+2a,t)-\rho^{-}(Na,t),~~~~~\label{me5}
\eea
where $x \in [2a,L]$, $L=(N-1)a$ and  $\delta_{x,x'}$ is the Kronecker-delta symbol. We set the left-hand side of Eqs.~\ref{me1}-\ref{me5} to zero in the steady state of the system at long times. We immediately find $\rho^{+}(a,t)=\rho_0+\delta \rho$ and $\rho^{-}(L+a,t)=\rho_0$ for all time. We can define the steady-state charge current flowing through the two ends of the lattice by applying the continuity equations. The currents flowing in and out of the lattice are given by $I_{\text{in}}=(ea/\tau)(\rho^{+}(a,t)-p\rho^{-}(2a,t))$ and $I_{\text{out}}=(ea/\tau)(p\rho^{+}(L)-\rho^{-}(L+a))$, respectively (see App.~\ref{Sec_1_1}).  For $N=4$, we find $\rho^{\pm}(x)$ by solving Eqs.~\ref{me1}-\ref{me5}, which gives
\begin{align}
I_{\text{in}}=I_{\text{out}}=\frac{(ea/\tau) \delta \rho- (1-p)I_g}{3-2p}.
\end{align}
These current values change to $((ea/\tau) \delta \rho+(1-p)I_g)/(3-2p)$ for the reverse bias. Thus, $I_{\text{in}}$ matches with the previous result of the circuit model in the limit $p \to 0$, when we identify group velocity $v_F=a/\tau$ and $e v_F \delta \rho/3=V/R$. The limit $p \to 0$ signals a large inter-conversion between left- and right-movers at each connection between S or D and the lattice. The above nonreciprocity in current flow disappears in the opposite limit of $p \to 1$ for such a short lattice with only one pair of S and D. Thus, the conversion between two species of charge carriers is the physical origin of nonreciprocal transport in this system in the presence of S and D.  The inter-conversion is equivalent to inelastic scattering of charge carriers as energy is not conserved in these processes and leads to resistance in such classical transport channel.

\begin{figure*}
\includegraphics[height=4.5cm,width=17.6cm]{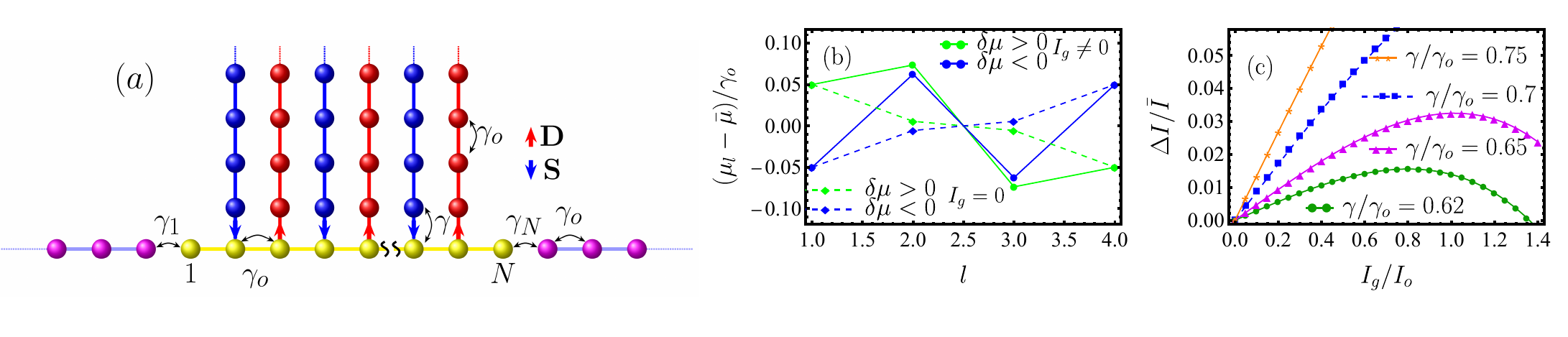}
\caption{(a) Schematic of an open quantum wire (yellow dots) connected to multiple sources (S's) and drains (D's) in the middle. \textcolor{black}{(b) Chemical potentials, $\mu_2$ and $\mu_3$, of S and D for a short quantum wire $(N=4)$ in the forward and reverse bias. Here $\bar\mu/\gamma_o=1.0$, $\gamma/\gamma_o=0.5$ and we set the non-zero value of $I_g$ as $I_g/I_o=0.3$.} (c) Relative current nonreciprocity, $\Delta I/\bar I$ with increasing $I_g$ beyond linear response regime of a short wire $(N=4,N_i=2)$. Here, $\Delta I=|I_1(\delta\mu<0)|-|I_1(\delta\mu>0)|$, $2\bar I=|I_1(\delta\mu>0)|+|I_1(\delta\mu<0)|$, and $\bar\mu/\gamma_o=1.52$.  In plots (b-c), we fix $\delta\mu/\gamma_o= 0.1$ and $\delta\mu/\gamma_o=- 0.1$ for forward and reverse bias, respectively, with $e=1$. }
\label{Fig1}
\end{figure*}

We next take an extended lattice with a large number $(N_i \gg 2)$ of S and D. We now solve a large set of linear coupled equations for $\rho^{\pm}(x)$ at the steady state of the lattice with the boundary conditions of $\rho^{+}(a,t)=\rho_0+\delta \rho$ and $\rho^{-}(L+a,t)=\rho_0$. These solutions for densities of both the species at odd and even sites are (see App.~\ref{Sec_1_3})
\bea
&\rho^-(x)=\frac{\tilde{\rho}-(2p-1)I_g/(2ev_F)}{p(1+(N-2)(1-p))},~x=3a,5a,\dots,L, \nonumber\\
&\rho^-(x)=\frac{\tilde{\rho}+N(1-p)I_g/(2ev_F)}{p(1+(N-2)(1-p))},~x=2a,4a,\dots,L-a,\nn\\
&\rho^{+}(x)=\rho^{-}(x)+\frac{\delta\rho+I_g/(2ev_F)}{1+(N-2)(1-p)},~x=2a,3a,\dots,L,\nn\\
&\tilde{\rho}=\rho_0+\big[(N-2)\rho_0+N\delta \rho-\frac{x}{a}(\delta \rho+\frac{I_g}{2ev_F})\big](1-p).~~
\eea
 We insert these densities in  the definition of $I_{\rm in}$  and $I_{\rm out}$ to find the steady-state charge current at the boundaries as (check App.~\ref{Sec_1_3})
\bea
I_{\rm in}=I_{\rm out}=\frac{ev_F \delta \rho-(1-p)(N/2-1)I_g}{1+(N-2)(1-p)}.\label{curME}
\eea
When the densities of the species at the two terminals are reversed, the value of $I_{\rm in}$ and $I_{\rm out}$ is not the same as before for a non-zero $I_g$ as long as $1-p>0$. Hence, we reaffirm that we need a finite inter-conversion of species to get nonreciprocity. We further define a length scale $l_c=a/(1-p)$, which we later identify as the scattering length that emerged due to the inter-conversion of species. Then, we can rewrite Eq.~\ref{curME} as
\bea
I_{\rm in}=I_{\rm out}=\frac{ev_F \delta \rho}{1+L/l_c}-\frac{I_g}{2}\bigg(1-\frac{1}{1+L/l_c}\bigg), \label{curME1}
\eea
where we assume $(N-2)a\approx L$ for $N\gg2$. Thus, we find the emergence of a finite $l_c$ is an essential entity for nonreciprocity as nonreciprocity vanishes for $l_c\to \infty$. Till now, we have discussed transport in classical channels without quantum coherence. It is also exciting how quantum coherence competes with nonreciprocity. For this, we now turn to charge transport in quantum channels. 

\section{Quantum Langevin equation approach }\label{Sec_QLEq}
 A full-fledged quantum transport analysis generalizing approach of Landauer and B{\"u}ttiker \cite{Mello2004, Dutta2005, Akkermans2007} requires quantum transport channel(s) connected to baths at the boundaries generating a voltage bias \cite{Bruus2004, Dhar_Ford-Kac-Mazur_PhysRevB2003, DharSen_BoundState_PhysRevB2006, DRoy_ElectronSelfConsistent_PhysRevB2007, Bondyopadhaya_Nonequilibrium_JStatPhys2022}. We consider a finite-length tight-binding (TB) wire of spinless fermions connected to two microscopic bath models at two ends. The baths are kept at different chemical potentials to create a voltage difference across the wire. We further model the sources and drains as microscopic baths whose chemical potentials are fixed self-consistently so that they inject or draw the required current $I_g$ to the transport channel \cite{Bolsterli1970, Visscher1975, Buttiker1985, Buttiker1986, D'Amato_DisorderChain_PhysRevB1990, Bonetto2004, Dhar_2006, DRoy_ElectronSelfConsistent_PhysRevB2007, Roy_2008, Roy_2008PRE, kulkarni_DensityMatrix_NewJphys2013, Saha_Quasiperiodic_PhysRevB2024}.  The chemical potentials of electrons can be perceived as a mathematical alternative to using electron densities in our master equation description. All the microscopic baths are modeled as semi-infinite TB chains, and all the baths are kept at the same temperature $T$ for simplicity of the analytical calculation. 

The Hamiltonian of the full system, depicted in Fig.~\ref{Fig1}(a), consisting of the wire plus all the baths, is given by $\hat{H}=\hat{H}_w+\sum_{l=1}^N(\hat{H}^l_B+\hat{V}^l)$, where (setting $\hslash=1$)  
\begin{align}
  &\hat{H}_w=-\gamma_o\sum_{l=1}^{N-1} ( \hat{c}_l^{\dagger}\hat{c}_{l+1}+\hat{c}^{\dagger}_{l+1}\hat{c}_{l})\label{wire_Hamiltonian},\\
  &\hat{H}^l_{B}= -\gamma_o\sum_{m=1}^{\infty} (\hat{b}_m^{\dagger}(l)\hat{b}_{m+1}(l)+\hat{b}^{\dagger}_{m+1}(l)\hat{b}_{m}(l)),\label{reservoirs'_Hamiltonian} \\
  & \hat{V}^l=-\gamma_l(\hat{b}^\dagger_1(l)\hat{c}_l+\hat{c}_l^\dagger \hat{b}_1(l)).\label{wire_reservoir_coupling}
\end{align}
Here, $\hat{H}_w$ describes a 1D quantum wire of $N$ sites. The operator $\hat{c}^\dagger_l$ ($\hat{c}_l$) creates (annihilates) a spinless fermion at site $l$, and it can hop to two neighboring sites with an amplitude $\gamma_o$. $\hat{H}_B^l$ denotes the Hamiltonian of the bath connected to site $l$ of the wire. Here, $\hat{b}^\dagger_m(l)$ ($\hat{b}_m(l)$) creates (annihilates) a spinless fermion at site $m$ inside a bath. We set the hopping amplitude within each TB bath also $\gamma_o$. The coupling Hamiltonian between $l^{th}$ site of the wire and a bath is $\hat{V}^l$, and $\gamma_l$ is the related coupling amplitude. The chemical potential of the left and right bath connected to site $l=1, N$, respectively, are fixed to be $\mu_{\text{L}}$ and $\mu_{\text{R}}$. The average and the difference of the left and right bath's chemical potentials are $\bar\mu=(\mu_{\text{L}}+\mu_{\text{R}})/2$ and $\delta\mu=\mu_{\text{L}}-\mu_{\text{R}}$, respectively. The chemical potential $\{\mu_l | l=2,3,..., N-1\}$ of the middle baths are determined self-consistently to get $I_g$ current injected (drawn) by the S (D) bath at an even (odd) site in the bulk of the wire. 
The self-consistency condition, required for the balance of the average S and D currents, does not fix the phase and energy of individual electrons exchanged with the baths. As a result, the side reservoirs effectively emulate the essence of inelastic scatterers within the wire. We further set $\gamma_l=\gamma$ for $l=2,.., N-1$ and $\gamma_o=1$, allowing all the coupling and hopping amplitudes to be chosen in the units of $\gamma_o$.

Let us assume that the baths are connected to the wire in the remote past $t_o \to -\infty$. Following the method in \cite{DharSen_BoundState_PhysRevB2006, DRoy_ElectronSelfConsistent_PhysRevB2007,Bondyopadhaya_Nonequilibrium_JStatPhys2022}, we get generalized quantum Langevin equations for the wire operators for site $l=1,2,3\dots, N$ as (see App.~\ref{Sec_2})
\begin{align}
    i\frac{\partial \hat{c}_l}{\partial t}=-(\hat{c}_{l+1}+\hat{c}_{l-1})+\int_{-\infty}^{t} d\tau\: \Sigma_l^{+}(t-\tau)\hat{c}_l(\tau)+\hat{\eta}_l(t),\label{QLE_wire}
\end{align}
where the noise $\hat{\eta}_l(t)=-i\gamma_l\sum_{m}{g}_{1,m}^{+}(l,t-t_o) \hat{b}_{m}(l,t_o)$ and the self-energy correction ${\Sigma}_l^{+}(t)=\gamma_l^2 {g}_{1,1}^{+}(l,t)$ generate fluctuation and dissipation in the wire due to its coupling with the bath at $l^{th}$ site. The open boundary condition on the wire gives $\hat{c}_0=\hat{c}_{N+1}=0$. \textcolor{black}{ The elements ${g}_{m,m'}^{+}(l,t)$ are obtained using the green's function operator of the bath at $l^{th}$ site:}
{\color{black}\begin{align}
\hat{g}^{+}(l,t)=-i\theta(t)  e^{-i\hat{H}^l_{B}t},
\end{align}}
where  $\theta(t)$ represents the Heaviside step function. 
\begin{figure*}
\includegraphics[height=4.6cm,width=17.5cm]{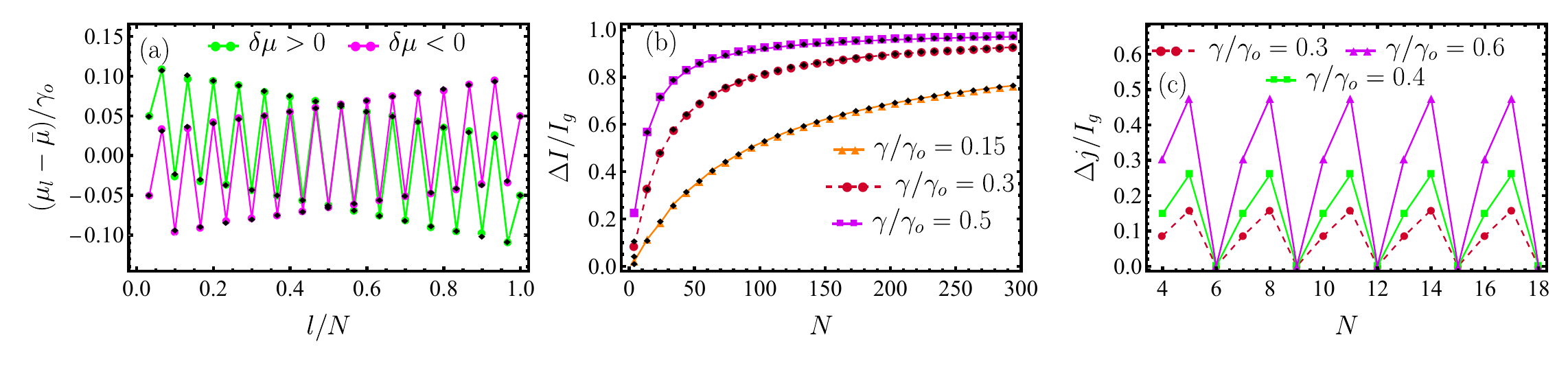}
\caption{\textcolor{black}{(a) \textcolor{black}{Chemical} potential profile in a long quantum wire $(N=30)$ connected to $N_i/2=N/2-1$ number of S-D dimers. Parameters are $\bar\mu/\gamma_o=1.0$, $I_g/I_o=0.3$, and $\gamma/\gamma_o=0.5$. Black dots represents the corresponding analytical points .} (b) Variation of current nonreciprocity $\Delta I=|I_1(\delta\mu<0)|-|I_1(\delta\mu>0)|$, with $N$ in a long wire with $N_i/2=N/2-1$ number of S-D dimers. Here, $I_g/I_o=0.05$, and $\bar\mu/\gamma_o=1.0$. The black dots represent the corresponding analytical points obtained using $2\delta I_{N}$ in Eq.~\ref{multiple SD current}. (c) Periodic variation of current nonreciprocity, $\Delta j$ with $N$ in a quantum wire with only one S-D dimer ($N_i=2$). Here, $I_g/I_o=0.1$, and $\bar\mu/\gamma_o=1.0$. In all the plots (a-c), we fix \textcolor{black}{$\delta\mu/\gamma_o=\pm 0.1$} and $e=1$. }
\label{Fig2}
\end{figure*}
When the wire attains a steady state at a long time, we can write the steady-state solution of Eq.~\ref{QLE_wire} in the frequency domain \cite{DharSen_BoundState_PhysRevB2006} as $\hat{{c}}_{l}(\omega)=\sum_{l'=1}^{N} {G}^{+}_{ll'}(\omega)\hat{{\eta}}_{l'}(\omega)$, where $\hat{{c}}_l(\omega)$, $\hat{{\eta}}_l(\omega)$ are the Fourier domain operators.  \textcolor{black}{The retarded Green's function elements  $G^{+}_{ll'}(\omega)$ of the wire are (App.~\ref{Sec_2_1})}
\begin{align}
\mathcolor{black}{{G}^{+}_{ll'}(\omega)= \langle \varphi|\hat{c}_{l}[\omega-\hat{H}_w-\hat{{\Sigma}}^+(\omega)]^{-1} \hat{c}^\dagger_{l'}|\varphi\rangle,}
\end{align}
where \textcolor{black}{  $|\varphi\rangle$ represents the vacuum mode of the full system.} The elements of the operator $\hat{{\Sigma}}^+(\omega)$ contains frequency domain self-energy corrections  as ${\Sigma}^+_{ll'}(\omega)={\Sigma}_l^+(\omega)\delta_{ll'}$.
The noise properties in the frequency domain are the following: $\langle \hat{\eta}_l(\omega) \rangle=0$ and 
\begin{align}
\langle\hat{\eta}_l^\dagger(\omega) \hat{\eta}_{l'}(\omega')\rangle =\gamma_l^2 f_l(\omega) \rho_l(\omega)\delta(\omega-\omega')\delta_{ll'},
\end{align} where $f_l(\omega)=1/(e^{(\omega-\mu_l)/K_B T}+1)$ is the Fermi distribution function and $\rho_l(\omega)=\sqrt{4-\omega^2}/(2\pi)$ is the local density of states at the first site of the $l^{th}$ bath. The averaging is performed over the grand-canonical distribution of the baths. We apply the continuity equations to define two kinds of charge currents in the system: (i) $I_l=ie\gamma_l \langle \hat{c}_l^\dagger \hat{b}_{1}(l)-\hat{b}_{1}^\dagger(l) \hat{c}_l\rangle$, which is the current flowing from the bath at $l^{th}$ site to that site of the wire, and (ii) $J_l=ie\langle \hat{c}_{l+1}^\dagger \hat{c}_l-\hat{c}_l^\dagger \hat{c}_{l+1} \rangle$ for $l=1,2,..., N-1$ denotes the current flowing from site $l$ to site $l+1$ inside the wire. To perform analytical calculation, we additionally assume a linear-response regime with small chemical potential biases about $\bar\mu$: $\delta\mu \ll \bar\mu$ and $\delta\mu_l=\mu_l-\bar\mu \ll \bar\mu$ for $l=2,3,..., N-1$. This simplifies the average current expressions in the steady state which are provided in App.~\ref{Sec_2_2}. We find $\mu_l$ of the middle baths from the self-consistency condition that $I_l=(-1)^{l}I_g$ for $l=2,3,\dots,N-1$.
We further set $\gamma_1/\gamma_o=\gamma_{N}/\gamma_o=1$, which leads to transparent (reflection-less) contacts for the two ends of the wire \cite{Dhar_Ford-Kac-Mazur_PhysRevB2003}. We denote $I_o=e\delta\mu/(2\pi)$ as the corresponding reflection-less ballistic current within the wire when $\gamma=0$.
\subsection{Short wire}
We first consider the case with $N=4$, where the wire contains a single S-D dimer. The chemical potentials of the middle baths are obtained as $\mu_2=\bar\mu+\Delta_\mu$ and $\mu_3=\bar\mu-\Delta_\mu$ (Fig.\ref{Fig1}(b)) (see App.~\ref{Sec_2_3} for the derivation). By expanding expanding $\Delta_{\mu}$ in the powers of $\gamma^2$, we get 
\begin{align}
    \Delta_{\mu}=\pi I_g/(e\gamma^2)+\pi \bar\mu^2  I_g/(2e)+\mathcal{O}(\gamma^2).\label{Delta_muEq}
\end{align}
 Thus, $I_g$ cannot be arbitrarily large for a small $\gamma$ for the system to be in a linear-response regime, e.g., $\Delta_{\mu} \leq \delta\mu$, which restricts $I_g \leq 2\gamma^2I_o$ when $\gamma/\gamma_o \ll 1$. \textcolor{black}{From Fig.~\ref{Fig1}(b), we observe that $\mu_2 $ and $\mu_3$ do not symmetrically interchange their values when we reverse the bias across the wire. This leads to nonreciprocity in electrical transport.} 

At long-time steady state of the wire, the charge current at the left and right boundaries for a non-zero $\delta\mu$ are related as $I_{1}=-I_{4}$, which can  be expressed as $I_1=G_{4}(\delta \mu/e) -\delta I$ (see App.~\ref{Sec_2_3}), where 
\begin{align}
 &G_4 \approx \frac{e^2}{2\pi}(1-\gamma^2),~ \delta I \approx \gamma^2(2-\bar{\mu}^2)\frac{I_g}{2},\label{G4_Eq}
\end{align}
for $\gamma/\gamma_o \ll 1$. Thus, the magnitude of $I_1$ for the forward bias ($\delta\mu>0$) is different from that for the reverse bias ($\delta\mu<0$) when $I_g \ne 0$. This leads to nonreciprocal charge transport in the wire in the presence of a pair of source and drain, as discussed earlier for the two previous models. The value for the nonreciprocity once again becomes $2I_g/3$ when $\gamma^2=(\bar\mu^2-1+\sqrt{5-2\bar\mu^2+\bar\mu^4})/2$ in the non-perturbative regime (check App.~\ref{Sec_2_3}). The relative nonreciprocity in the linear response regime is given by $2e\delta I/(G_4\delta \mu)$, which grows linearly with $I_g$. Interestingly, the relative nonreciprocity behaves differently at larger $I_g$ in the nonlinear response regime (see App.~\ref{Sec_2_3}) as shown in Fig.~\ref{Fig1}(c).

\subsection{Long wire}

 Next, we discuss charge transport in a longer wire ($N \gg 4$) with $(N/2-1)$ S-D dimers connected to the middle sites. 
For $\gamma<1$ and $N\gg 4$, we self-consistently determine $\mu_l$ of the S's and D's in the middle as (check App.~\ref{Sec_2_4})
\begin{align}
       &\mu_l=\mu_{\text{L}}-\phi+\Big[\frac{l_c}{2\sigma}+(-1)^l\nabla\Big]\frac{eI_g}{2} -2\frac{\phi}{l_c} (l-2), \label{chempot__multipleSD} \\
    &\phi=\frac{\delta\mu+eI_g l_c/(2\sigma)}{2[1+(N-3)/l_c]},~~\text{and}~~ \nabla= \frac{\coth^2{\alpha_R}}{2 \sigma},
\end{align}
for $l=2,3,...,N-1$. Here, we define the scattering length $l_c = 1/\alpha_R$ and the electrical conductivity  $\sigma=e^2\sin^2{\alpha_I}\coth{\alpha_R}/(2\pi |\sinh{\alpha}|^2)$. Further, $\alpha_R$ and $\alpha_I$ are, respectively, the real and imaginary parts of the complex-valued function $\alpha=\log_e{[\mathcal{C}+\sqrt{\mathcal{C}^2-1}]}$ with $\alpha_R>0$ and $2\mathcal{C}=\bar\mu-\gamma^2\big(\bar\mu-i\sqrt{4-\bar\mu^2}\big)/2$.  
We show $\mu_l$ over the wire sites $l$ for $I_g \neq 0$ in Fig.~\ref{Fig2}(a), where we depict an excellent match between the analytical expression of $\mu_l$ (Eq.~\ref{chempot__multipleSD}) with exact numerical results for a fixed $N$ and  $\gamma$. Due to local injection and extraction of particles by the S and D, the charge current $J_l$ at an even and an odd site of the wire are not the same, and we call them by $J_{\text{S}}$ and $J_{\text{D}}$, respectively. Applying $\mu_l$ from Eq.~\ref{chempot__multipleSD} in $J_l$, we find $J_{\text{D}}$ and $J_{\text{S}}$ for $N\gg 4$ as $J_{\text{D}}=G_{N}(\delta\mu/e) -\delta J$ and $J_{\text{S}}=J_{\text{D}}+I_g$, where 
\begin{align}
 G_{N}=\frac{\sigma }{l_c+N-3},~\delta J=\frac{I_g}{2}\bigg(1-\frac{l_c}{l_c+N-3}\bigg).\label{multiple SD current}
\end{align}
We observe $J_{\text{D}}=I_1=-I_N$ from the continuity equation at steady state. Thus, the particle current at the left and right junction between the wire and respective baths are $J_{\text{D}}$. We further observe that $I_1$ or $I_N$ is again different  for forward $(\delta \mu>0)$ and reverse bias $(\delta \mu<0)$. The value of nonreciprocity in charge current is $2 \delta J$, which matches that obtained within the master equation analysis. Coupling of S and D baths to all middle sites of the wire generates inelastic scattering for the charge carriers, which leads to loss of coherence and energy dissipation (Joule heating  \textcolor{black}{in App.~\ref{Sec_2_4}}). These scatterings give rise to a large resistance in the wire against particle transport, and the resistance itself depends on the length of the wire. Thus, we find an emergent nonreciprocal Drude-type transport in the quantum wire in the presence of S and D baths to all middle sites. We can further relate the inelastic scattering due to the S and D baths in quantum channels to the inter-conversion of species in classical channels from the similarities of Eq.~\ref{multiple SD current} and Eq.~\ref{curME1}, when we identify $\sigma (\delta \mu/e)$ with $e v_F \delta \rho \:l_c$.  
 
The emergence of $l_c$ due to the coupling of the wire with middle baths leads to two regimes of quantum transport depending on $N$ and $l_c$. When $N \gg l_c$ for a long wire with a finite $\gamma$, we observe that $\delta J$ can be approximated as $I_g/2$, which gives $J_{\text{D}}=G_{N}(\delta\mu/e) -I_g/2, J_{\text{S}}=G_{N}(\delta\mu/e)+I_g/2$ and the nonreciprocity $2\delta J=I_g$. On the contrary, when $N \ll l_c$, we get the nonreciprocity $2\delta J \approx \gamma^2(N-3)I_g/2$ by approximating $l_c=\alpha_R^{-1} \approx 2/\gamma^2, \sigma \approx e^2l_c/(2\pi)$ and $\nabla \approx \pi l_c/e^2$ for $\gamma \ll 1$. In Fig.~\ref{Fig2}(b), we display two different regimes of nonreciprocity by plotting $2 \delta J$ with $N$. It shows a linear growth of $2 \delta J$ with $N$ for shorter $N$ $( \ll l_c )$ and saturation to $I_g$ at longer $N$.

\subsection{Quantum coherence}\label{Sec_QCoh}
 The coherence in charge transport through the long quantum wire is severely suppressed when all middle sites of the long wire are connected to S and D baths. However, it plays a vital role in transport when we connect some or only two middle sites to the S and D baths. Next, we analyze how coherence controls nonreciprocity in a quantum wire. We take a quantum wire with $N$ sites, where an S and a D bath (e.g., $N_i=2$) are connected to the sites  $l=2$ and $N-1$, respectively. All the $(N-4)$ middle sites for $l=3,4,\dots, N-2$ are free and form ballistic channels for charge transport. The wire is again under a finite bias $\delta \mu$ from the left and right boundary. The chemical potential of S and D baths can be derived from the requirement of $I_g$ current injected in and extracted from the wire in the steady state. 
\begin{figure}[h!]
\includegraphics[height=3.1cm,width=8.2cm]{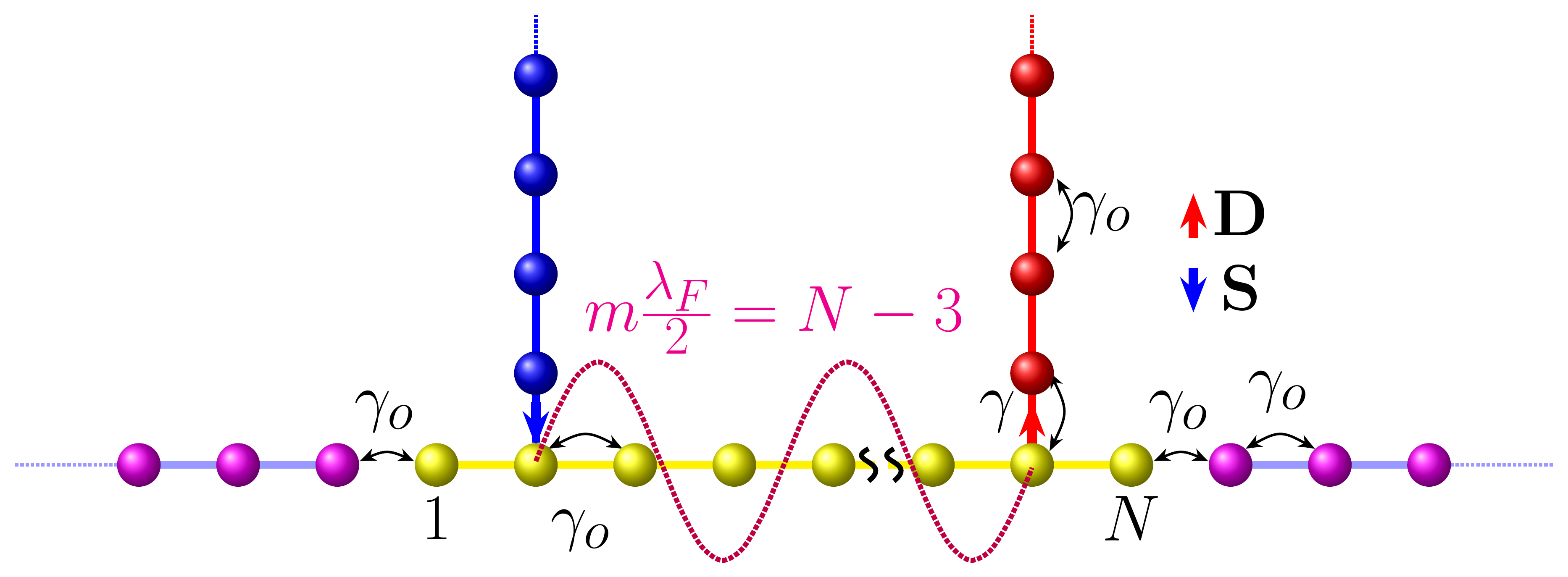}
\caption{\textcolor{black}{Quantum coherence in charge transport. The nonreciprocity in electrical current goes to zero when the length $(N-3)$ becomes integer $(m)$ times $\lambda_{F}/2$.} }
\label{FigCoherence}
\end{figure} 
 Applying those chemical potentials, we find the current $I_1, I_N$ at the boundaries, showing nonreciprocity under reversal of voltage bias $\delta \mu$. The nonreciprocity in charge current for this case is given by $\Delta j= 2(\Xi-1)I_g/(1+2\gamma^2+\Xi)$, where (\textcolor{black}{check App.~\ref{Appendix_One S-D in a long ballistic quantum wire}})
\begin{align}
\Xi=\frac{\big|(2\mathcal{A}\mathcal{C}-1)\sin{(N-3)k_F}-\mathcal{A}\sin{(N-4})k_F\big|^2}{\sin^2{k_F}},
\end{align}
\textcolor{black}{$\mathcal{A}=\bar\mu/2+i\sqrt{1-\bar\mu^2/4}$}, and the Fermi momentum $k_F=\cos^{-1}({\bar\mu/2})$, which gives Fermi wavelength $\lambda_F=2\pi/\cos^{-1}({\bar{\mu}/2})$. We show $\Delta j$ with varying $N$ in Fig.~\ref{Fig2}(c) for $\bar{\mu}=1$ giving $\lambda_F=6$. We observe $\Delta j$ oscillates with $N$ for different $\gamma$, displaying an interplay between quantum coherence and nonreciprocity. Further, $\Delta j$ vanishes when the length ($N-3$) between S and D baths is an integer multiple of $\lambda_F/2$ \textcolor{black}{as shown in Fig.\ref{FigCoherence}}. Then, the \textcolor{black}{chemical} potential of S and D remains the same for both forward and reverse voltage bias. Such oscillations and zeros of nonreciprocity do not appear in a similar set-up within the master equation analysis for classical transport channels (\textcolor{black}{see App.~\ref{Appendix_One S-D in a long ballistic classical wire}}).  

\section{Conclusions} 
 The nonreciprocal transport mechanism in our models has similarity to other magnet-free rectifications using an active drive to break reciprocity, e.g., spatiotemporal modulations \cite{Yu2009}. While the internal characteristics of the wire with balanced current S's and D's remain unaltered for both forward and reverse bias (e.g., the resistive circuits in Fig.~\ref{Fig0}), the \textcolor{black}{chemical} potentials of these S's and D's are required to readjust for different bias to maintain a fixed $I_g$ in our quantum analysis. Such readjustment challenges the defining feature of rectifiers or isolators. It also raises questions on conditions for balanced gain and loss in many experiments demonstrating nonreciprocity with effective $\mathcal{PT}$ symmetric media, where no extra care is taken for such balanced gain and loss for forward and reverse bias. Further, our analysis shows two necessary requirements for nonreciprocity are: individual breaking of parity and time-reversal symmetry in the models, which are valid for other rectification mechanisms, including the Faraday isolators \cite{Stadler2014} and nonlinear rectifiers \cite{RoyPRB2010}. Thus, our results suggest that these systems with broken parity and time-reversal symmetry show an arrow of space manifested through nonreciprocal transport. 
Finally, it would be exciting to extend our current analysis with balanced loss and gain to topological wires, e.g., the Su–Schrieffer–Heeger (SSH) chains \cite{Nair_DissipativeSSH_PhysRevB2023, Min_ManybodyBathInducedTopology_arxiv2024} and the Majorana wires \cite{Bondyopadhaya_Nonequilibrium_JStatPhys2022,Antipov_MajoranaNanonwires_PhysRevX2018,Pagano_IronSuperconductingNanowire_2020NanoMat}. {\color{black} Another extension can be to explore the role of long-range hopping and the arrangement of these balanced losses and gains to enhance the current nonreciprocity. }

\section*{Appendix}
\appendix

\textcolor{black}{The Appendix is organized into four sections as follows. The App.~\ref{SEC_1} contains details of the master equation approach for the charge transport in a lattice model. We derived the expression for the incoming and the outgoing electrical currents using the charge continuity equation in App.~\ref{Sec_1_1}. The steady-state solutions of the density fields of particles for single and multiple S-D dimers are discussed in App.~\ref{Sec_1_2} and App.~\ref{Sec_1_3}, respectively. We further provide a continuum limit description of the lattice model in the steady states in App.~\ref{Sec_1_4}.}

\textcolor{black}{We provide the details of the charge transport through a quantum wire using the quantum Langevin equation approach in App.~\ref{Sec_2}. The steady-state solutions, noise properties and electrical current expressions are discussed in App.~\ref{Sec_2_1} and App.~\ref{Sec_2_2}. The derivation of the analytical expressions for the charge currents for  single and multiple S-D dimers are given in App.~\ref{Sec_2_3} and  App.~\ref{Sec_2_4}, respectively. Further, we discuss the competition  between the coherence effect and the nonreciprocity in the wire due to imperfect contacts in App.~\ref{Contact_scattering_appendix}.}

\textcolor{black}{The App.\ref{Sec_3} contains details of charge transport in an extended wire (or a lattice) with one S and one D. In App.~\ref{Appendix_One S-D in a long ballistic quantum wire}, we provide detailed calculations for the derivation of \textcolor{black}{chemical} potentials of the S-D reservoirs and charge currents, which is presented in Sec. \ref{Sec_QCoh}. In App.~\ref{Appendix_One S-D in a long ballistic classical wire}, we provide a similar set-up in a classical transport channel and discuss the nonreciprocal transport using the master equation description. }

\textcolor{black}{Finally, in App.~\ref{resistive model}, we discuss nonreciprocal charge transport in a resistive circuit model with multiple S-D dimers.} 
\section{Master equation approach for charge transport in a 1-D lattice}\label{SEC_1}
 In this section, we give details of the master equation approach
described in Sec.~\ref{Sec_MEq} of the main text to study charge transport in a 1D lattice coupled to $N_i/2=N/2-1$  pairs of Ss and Ds. The length of the lattice is denoted by $L=(N-1)a$. Using Eqs.~\ref{me1}-\ref{me5} in the main text, we write the following steady-state equations for the density fields of the right- and left-moving particles: 
\begin{align}
    &\rho^{+}(a)=\rho^{+}(0),~~\rho^{-}(a)=p\rho^{-}(2a),\label{Eq1}\\
&[p-(p-1)\delta_{x,2a}]\rho^{+}(x- a)-p\rho^{+}(x)+(1-p)\nonumber\\
&~~~~~~~~~~~~~~~~~~~~~~~[\rho^{-}(x)-\rho^{+}(x)]+(-1)^{\frac{x}{a}}\frac{I_g }{2e v_F}=0,\label{Eq2}
\end{align}
\begin{align}
&[p-(p-1)\delta_{x,(N-1)a}]\rho^{-}(x+ a)-p\rho^{-}(x)+(1-p)\nonumber\\&~~~~~~~~~~~~~~~~~~~~~~~[\rho^{+}(x)-\rho^{-}(x)]+(-1)^{\frac{x}{a}}\frac{I_g }{2ev_F}=0,\label{Eq3}\\
&p\rho^{+}(L)=\rho^{+}(L+a),~~\rho^{-}(L+a)=\rho^{-}(L+2a),\label{Eq4}
\end{align}
where $x\in[2a,L]$, and we denote  $v_F=a/\tau$ as the group velocity of the mobile charges. The boundary conditions for the above equations are mentioned earlier before Eq.~\ref{me1}. 
 
\subsection{Boundary and bulk currents }\label{Sec_1_1}
Here, we derive various current formulae using the continuity equations for the charge densities.
The total charge density, at any location $x$ (and time $t$) is $e[\rho^{+}(x,t)+\rho^{-}(x,t)] $. We first focus on the left boundary of the lattice. Utilizing the master equations in Eqs.~\ref{me1}-\ref{me5}, we get:
 \begin{align}
   e\frac{\partial}{\partial t} \Big[\rho^{+}(a,t)+\rho^{-}(a&,t)\Big]  =ev_F\bigg[\frac{\rho^{+}(0,t)-\rho^{-}(a,t)}{a}\bigg]\nonumber\\
    &-ev_F\bigg[\frac{\rho^{+}(a,t)-p\rho^{-}(2a,t)}{a}\bigg],
 \end{align}
where, we use $\frac{\partial}{\partial t} \rho^{\pm}(x,t)\equiv [\rho^{\pm}(x,t+\tau)-\rho^{\pm}(x,t)]/\tau$. Now, we can identify the charge current flowing into the lattice as follows:
$I_{\text{in}}=ev_F[\rho^{+}(a,t)-p\rho^{-}(2a,t)]$. Similarly, at the right boundary, we get :
\begin{align}
   & e\frac{\partial}{\partial t} \Big[\rho^{+}(L+a,t)+\rho^{-}(L+a,t)\Big]=ev_F\bigg[\frac{p\rho^{+}(L,t)}{a}-\nonumber\\
    &~~~\frac{\rho^{-}(L+a,t)}{a}\bigg]-ev_F\bigg[\frac{\rho^{+}(L+a,t)-\rho^{-}(L+2a,t)}{a}\bigg].
 \end{align}
Thus, the charge current flowing out of the lattice is defined using $I_{\text{out}}:=ev_F[p\rho^{+}(L,t)-\rho^{-}(L+a,t)]$. Next, we obtain the continuity equation for the other sites $x\in\{3a,4a,..., L-a\}$ in the bulk:
\begin{align}
    &e\frac{\partial}{\partial t} \Big[\rho^{+}(x,t)+\rho^{-}(x,t)\Big]=epv_F\frac{\rho^{+}(x-a,t)-\rho^{-}(x,t)}{a}\nonumber\\
    &~~~~~~~~~~~-epv_F\frac{\rho^{+}(x,t)-\rho^{-}(x+a,t)}{a}+(-1)^{\frac{x}{a}}\frac{I_g}{a}.
 \end{align}
This gives the current flowing inside the lattice from site $x$ to $x+a$ as  $J(x):=  epv_F [\rho^{+}(x,t)-\rho^{-}(x+a,t)]$. For odd (even) sites $x$, $J(x)$ gives the inter-(intra-) bond current within the lattice.

\subsection{ Single S-D dimer  }\label{Sec_1_2}
 Let us first consider the case of $N=4$, with a single pair of S and D coupled at sites $x=2a$ and $x=3a$, respectively. The linear set of equations in Eqs.~\ref{Eq1}-\ref{Eq4} can be inverted easily, and we obtain the following solutions ($p \neq 0$) for the density fields in the steady state:
\begin{align}
   &\rho^{-}(a)=\frac{2(1-p) [\frac{I_g}{2ev_F}+\delta\rho]}{3-2p}+\rho_o,\\
   &\rho^{+}(2a)=\frac{(2-p)[\frac{I_g}{2ev_F}+\delta\rho]+(3-2p)\rho_o}{p(3-2p)},\\
      &\rho^{-}(2a)=\frac{\rho_o +2(1-p) [\frac{I_g}{2ev_F}+\delta\rho+\rho_o]}{p(3-2p)},\\
      &\rho^{+}(3a)=\frac{(p-1)\frac{I_g}{ev_F}+\delta\rho+(3-2p)\rho_o}{p(3-2p)},\\
    &\rho^{-}(3a)=\frac{(p-2)\frac{I_g}{2ev_F}+(1-p)\delta\rho+(3-2p)\rho_o}{p(3-2p)},\\
    &\rho^{+}(4a)=\frac{(p-1)\frac{I_g}{ev_F}+\delta\rho+(3-2p)\rho_o}{3-2p}.
\end{align}
Therefore, the current flowing in (out of) the lattice is
\begin{align}
    I_{\text{in}}=I_{\text{out}}=\frac{e v_F \delta\rho-(1-p) I_g }{3-2p},
\end{align}
which is given in Sec.~\ref{Sec_MEq} of the main text, and the current between the S-D bond is 
\begin{align}   J(2a)=\frac{e v_F \delta\rho+(2-p) I_g }{3-2p}.
\end{align}
The above solutions suggest that the density field at the middle sites diverges as $1/p$ if we take $p \to 0$ (large inter-conversion). On the other hand, the middle current $J(2a)$ in the limit $p \to 0$  remains finite, since $J(2a)\propto p[\rho^{+}(2a)-\rho^{-}(3a)]$.

\subsection{Multiple S-D dimers in a long lattice}\label{Sec_1_3}
Next, we follow an iterative procedure to find the density profile in an extended lattice ($N>4$). At the odd sites $x=3a,5a,\dots,(N-1)a$ in the middle of the lattice, we get:
\begin{align}
    &\rho^-(x)=\frac{\splitfrac{\rho_0+\big[(N-2)\rho_0+N\delta \rho-\frac{x}{a}(\delta \rho}{+\frac{I_g}{2ev_F})\big](1-p)-(2p-1)\frac{I_g}{2ev_F}}}{p[1+(N-2)(1-p)]},\\
&\rho^+(x)=\frac{\splitfrac{\rho_0+\big[(N-2)\rho_0+N\delta \rho-\frac{x}{a}(\delta \rho}{+\frac{I_g}{2ev_F})\big](1-p)+p\delta\rho+(1-p)\frac{I_g}{2ev_F}}}{p[1+(N-2)(1-p)]}.
\end{align}
Similarly, at the even sites $x=2a,4a,\dots,(N-2)a$, we get : 
\begin{align}
&\rho^-(x)=\frac{\splitfrac{\rho_0+\big[(N-2)\rho_0+N\delta \rho-\frac{x}{a}(\delta \rho}{+\frac{I_g}{2ev_F})\big](1-p)+N(1-p)\frac{I_g}{2ev_F}}}{p[1+(N-2)(1-p)]},\\
&\rho^+(x)=\frac{\splitfrac{\rho_0+\big[(N-2)\rho_0+N\delta \rho-\frac{x}{a}(\delta \rho}{+\frac{I_g}{2ev_F})\big](1-p)+p\delta\rho+[N-(N-1)p]\frac{I_g}{2ev_F}}}{p[1+(N-2)(1-p)]}.
\end{align}
The above expressions exclude solutions at the boundaries of the lattice, and they are given by
\begin{align}
 & \rho^{-}(a)=\frac{(N-2)(1-p)[\delta\rho+\frac{I_g}{2ev_F}]}{1+(N-2)(1-p)}+\rho_o,\\
&\rho^{+}(L+a)=\frac{(\rho_o+\delta\rho)+(N-2)(1-p)[\rho_o-\frac{I_g}{2ev_F}]}{1+(N-2)(1-p)}.
\end{align}
Utilizing the above density fields, we obtain the intra-bond currents in the lattice as follows :
\begin{align} J(x)=\frac{e v_F \delta\rho+I_g/2}{1+(N-2)(1-p)}+\frac{I_g}{2},\label{intrabond_lattice}
\end{align}
where $x=2a,4a,...,(N-2)a$. The inter-bond currents in the lattice are obtained as  $J(x)=\frac{e v_F \delta\rho-(1-p)(N/2-1)I_g}{1+(N-2)(1-p)}=I_{\text{in}}=I_{\text{out}}$,
where $x=3a,5a,...,(N-3)a$, which are the currents in Eq.~\ref{curME}.

\subsection{Multiple S-D dimers in the continuous limit}\label{Sec_1_4}
We now consider the continuum limit ($a\to 0$) of the extended lattice model with multiple S-D dimers. We take a large number of dimers ($N_i/2 \gg 1$) and use a bipartite lattice description to make further progress. We introduce $X_e=2ma$ for $m=1,.., N/2$, and $X_o=(2m-1)a$ for $m=1,.., N/2$ to denote the positions of even and odd sites, respectively.  Our goal is to rewrite Eqs.~\ref{Eq1}-\ref{Eq4} with the help of the redefined coordinates, $X_e$ and  $X_o$.
To facilitate such a treatment, we consider the limit $p \to 1$ on all the hopping probabilities, but  
 retain a finite conversion rate  $\lim[(1-p)/\tau]\neq 0$ at all the middle sites of the lattice 
  where the S and D are coupled. 
 With these assumptions, the steady-state equations at the middle sites become:
\begin{align}
   &\rho^{\pm}(X_e \mp a)=\rho^{\pm}(X_e)\pm \frac{a }{l_c}
   \big[\rho^{+}(X_e)-\rho^{-}(X_e)\big]-\frac{I_g }{2e v_F},\label{S20}\\
   &\rho^{\pm}(X_o \mp a)=\rho^{\pm}(X_o)\pm \frac{a }{l_c}\big[\rho^{+}(X_o)-\rho^{-}(X_o)\big]
   +\frac{I_g}{2e v_F},\label{S21}
\end{align}
 where $X_e\in\{2a,4a,...,L-a\}$, and $X_o\in\{3a,5a,...,L\}$. Here, $ l_c=a/(1-p)$ denotes the scattering length scale in the problem.
 We aim to decouple the equations for the density fields on the even sublattice from those on the odd sublattice. The resulting recurrence equations will no longer contain the inhomogeneous driving term $I_g$, and we can take the continuum limit $a \to 0$ (along with $\tau \to 0$) to describe the density field of the particles on separate sublattices. Here, $I_g$ enters the description through the modified boundary conditions, as we see in the following discussions.

{\it{Equation for  $\rho^{+}(X_o)$} }:  Utilizing the above equations on the even sublattice, we get: 
\begin{align}
    \rho^{+}(X_e)-\rho^{-}(X_e)=\frac{ \rho^{+}(X_e-a)-\rho^{-}(X_e+a)}{1+2(a/l_c)} ,\label{S22}
\end{align}
where $X_e=2a,4a,...,L-a$. We focus on the steady-state equations for the right-moving charges at the odd sites.
If $X_o$ is an odd site, then $(X_o-a)$ will be positioned on the even sublattice. With the help of Eq.~\ref{S22}, the equations at  even sites $X_e=X_o-a$ for the right-moving particles  become:
\begin{align}
     &\rho^{+}(X_o-a)=\rho^{+}(X_o-2a)-
   \frac{a}{2a+l_c}\big[ \rho^{+}(X_o-2a)-\nonumber\\
   &~~~~~~~~~~~~~~\rho^{-}(X_o)\big]+\frac{I_g}{2e v_F},~\text{for}~X_o=3a,5a,..., L.\label{S23}
\end{align}
By substituting the above expression for $\rho^{+}(X_o-a)$ into Eq.~\ref{S21}, the inhomogeneous term $I_g/(2ev_F)$ cancels out and we arrive at the relations:
\begin{align}
    &\rho^{+}(X_o-\delta X_o)-\rho^{+}(X_o)-\frac{a\big[\rho^{+}(X_o-\delta X_e)-\rho^{-}(X_o)\big]}{2a+l_c}\nonumber\\
    &~~~~~-\frac{a}{l_c}\big[\rho^{+}(X_o)-\rho^{-}(X_o)\big]=0,~X_o=3a,5a,..., L. \label{S24}
\end{align}
In the above recurrence equation, the density fields on the odd sublattice are decoupled from those on the even sublattice. Here, $\delta X_o=2a$ represents the lattice constant of the odd sublattice. We divide both sides of Eq.~\ref{S24} by $\delta X_o$  and assume the continuum limit $a\to 0$. This leads to the following first-order differential equation: 
\begin{align}
  & \frac{\partial \rho^{+}(X_o)}{\partial X_o}+ \frac{1 }{l_c}\big[\rho^{+}(X_o)-\rho^{-}(X_o)\big]=0.
\end{align}
The variable $X_e$ now assumes continuous values within the interval $(0,L]$, and the boundary condition for the equation is fixed by $\rho^{+}(X_o=0^{+})=\rho_o+\delta\rho$. Next, we derive a similar equation for the left-moving particles.

{\it{Equation for  $\rho^{-}(X_o)$} }: Following the method of the previous paragraph, we now use $X_e=X_o+a$ and rewrite Eq.~\ref{S20} for the  left-moving particles as follows :
\begin{align}
     &\rho^{-}(X_o+a)=\rho^{-}(X_o+2a)+
   \frac{a}{2a+l_c}\big[\rho^{+}(X_o)\nonumber\\
   &~~~-\rho^{-}(X_o+2a)\big]+\frac{I_g}{2ev_F},~\text{for}~X_o=a,3a,..., L-2a.\label{S26}
\end{align}
Substituting this expression into Eq.~\ref{S21} gives the following  decoupled  equations :
\begin{align}
    &\rho^{-}(X_o+\delta X_o)-\rho^{-}(X_o)+\frac{a \big[\rho^{+}(X_o)-\rho^{-}(X_o+\delta X_o)\big]}{2a+l_c}+\nonumber\\
    &~~~~\frac{a}{l_c}\big[\rho^{+}(X_o)-\rho^{-}(X_o)\big]=0,~X_o=3a,..., L-2a. \label{S27}
\end{align}
The boundary value for the above recurrence relations is fixed at the rightmost site on the even sublattice, i.e., $X_o=L$. Using the steady-state equation at $x=L$, we get  
\begin{align}
    \rho^{-}(L)&=\rho^{-}(L+a)-\frac{I_g}{2ev_F}+(1-p)\rho^{+}(L)\nonumber\\
    &=\rho_o-\frac{I_g}{2ev_F}+a\bigg[\frac{\rho^{+}(L)}{l_c}\bigg]\nonumber  \\
    &=\rho_o-\frac{I_g}{2ev_F}+\mathcal{O}(a),\label{S28}
\end{align}
where the last term in the above expression is of the order of $a$. We divide both sides of Eq.~\ref{S27} by $\delta X_o$ and take the limit $a \to 0$. This yields :
\begin{align}
  & \frac{\partial \rho^{-}(X_o)}{\partial X_o}+ \frac{1  }{l_c}\big[\rho^{+}(X_o)-\rho^{-}(X_o)\big]=0.\label{S29}
\end{align}
Once again, $X_o$ in the above equation is a continuous parameter within the interval $(0, L]$ and the boundary value for Eq.~\ref{S29} becomes $\rho^{-}(X_o=L) \approx \rho_o-I_g/(2e v_F)$, after neglecting all the terms in $\rho^{-}(X_o=L)$ that are of the order of $a$.

{\it{Equation for  $\rho^{\pm}(X_e)$} }: One can check that the recurrence equations for the density field of the right-moving particles at even sites take the same form as Eq.~\ref{S24}, where $X_o$ is replaced by $X_e=4a,6a..., L-a$ and $\delta X_o$ is replaced by $\delta X_e=2a$. Here, the boundary value is fixed at the left-most site of the even sublattice, i.e., $X_e=2a$, which is given  by the following relation 
\begin{align}
    \rho^{+}(2a)&=\rho^{+}(a)+\frac{I_g}{2ev_F}+(1-p)\rho^{-}(2a)\nonumber\\
    &=\rho_o+\delta\rho+\frac{I_g}{2ev_F}+\mathcal{O}(a).\label{S28}
\end{align}
In a similar fashion, we find that $\rho^{-}(X_e)$ at odd sites satisfy Eq.~\ref{S27}, where we replace $X_o$ by $X_e=2a,4a,..., L-a$. At the boundary $X_e=L+a$, we have $\rho^{-}(X_e=L+a)=\rho_o$. Now, $X_e$ takes continuous values in the limit $a \to 0$ and we get:
\begin{align}
  &\frac{\partial \rho^{+}(X_e)}{\partial X_e}+ \frac{1 }{l_c}\big[\rho^{+}(X_e)-\rho^{-}(X_e)\big] =0,~X_e\in(0,L),\\
    &\frac{\partial \rho^{-}(X_e)}{\partial X_e}+ \frac{1}{l_c} \big[\rho^{+}(X_e)-\rho^{-}(X_e)\big]=0,~X_e\in(0,L],
\end{align}
where the boundary conditions for the above equations are $\rho^{+}(X_e=0^{+})\approx\rho_o+\delta\rho+I_g/(2ev_F)$ (after removing all terms of $\mathcal{O}(a)$ from this expression), and $\rho^{-}(X_e=L+0^{+})=\rho_o$, respectively. 

The steady-state density solutions for the middle sites in the lattice are as follows :
\begin{gather}
    \rho^{+}(X_o)= -\frac{C}{ l_c} X_o+\delta\rho+\rho_o,\label{S32}\\
    \rho^{-}(X_o)=\frac{C}{ l_c} (L-X_o)+\rho_o-\frac{I_g}{2ev_F}, \label{S33}\\
    \rho^{+}(X_e)=-\frac{C}{ l_c}X_e+\delta\rho+\rho_o+\frac{I_g}{2 ev_F}, \label{S34} \\
    \rho^{-}(X_e)= \frac{C}{l_c}(L-X_e)+\rho_o,\label{S35}
\end{gather}
where $C=[\delta\rho+I_g/(2ev_F)]/(1+L/ l_c)$.
Now, the density fields at the boundaries are obtained using the relations:
\begin{align}
    &\rho^{-}(X_o=a)=\lim_{\substack{p \to 1 \\a \to 0 }} p\rho^{-}(X_e=2a)=\rho_o+C \frac{L}{l_c},\\
    &\rho^{+}(X_e=L+a)=\lim_{\substack{p \to 1 \\ a \to 0 }} p\rho^{+}(X_o=L)=\delta\rho+\rho_o-C\frac{L}{ l_c}.
\end{align}
We observe that $\lim_{a\to 0}[\rho^{-}(X_o=a)-\rho^{-}(X_o=3a)]=I_g/(2ev_F)$, and $\lim_{a \to 0}[\rho^{+}(X_e=L-a)-\rho^{+}(X_e=L+a)]=I_g/(2ev_F)$. Therefore, the net density field suffers a discontinuity of $I_g/(2ev_F)$ at the right (left) boundary of the even (odd) sublattice. 
Using the full solutions, we derive the total particle densities at the middle sites as follows:
\begin{align}
  &\rho^{+}(X_e)+\rho^{-}(X_e)
  =2(\rho_o+\delta\rho) -\frac{2\delta\rho+I_g/(ev_F)}{ l_c+L}X_e\nonumber\\
  &~~~~~~~~~~~~~~~~~~~~~~-\frac{2\delta\rho-[I_g/(ev_F)](L/ l_c)}{2(1+L/ l_c)}+\frac{I_g}{2ev_F},\label{S39}\\
   &\rho^{+}(X_o)+\rho^{-}(X_o)=2(\rho_o+\delta\rho) 
  -\frac{2\delta\rho+I_g/(ev_F)}{ l_c+L}X_o\nonumber\\
  &~~~~~~~~~~~~~~~~~~~~~~-\frac{2\delta\rho-[I_g/(ev_F)](L/ l_c)}{2(1+L/ l_c)}-\frac{I_g}{2ev_F}.\label{S40}
\end{align}
The incoming and outgoing charge currents are
\begin{align}  I_{\text{in}}&=\lim_{\substack{p \to 1 \\ a\to 0 }}ev_F\Big[\rho^{+}(X_o=a)-p\rho^{-}(X_e=2a)\Big]\nonumber\\
    &=\frac{e v_F \delta\rho}{1+L/ l_c}-\frac{I_g}{2}\Bigg\{1-\frac{1}{1+L/ l_c}\Bigg\}=I_{\text{out}},\label{S41}
\end{align}
which are given in Eq.~\ref{curME1}.  We use current probes for the S's and D's in the master equation treatment and in the discussion of the resistive circuit model (Sec.~\ref{Sec_Intro} and  App.~\ref{resistive model}). Next, we discuss charge transport in the quantum regime using self-consistent voltage probes for the S's and D's.
\section{Quantum Langevin equation approach for charge transport in microscopic model}\label{Sec_2}
In this section, we provide details of the microscopic analysis of charge transport through a quantum channel (wire), which is discussed in Sec.\ref{Sec_QLEq}. The wire contains $N$ sites and is connected to microscopic baths. The full system (the wire and all the baths) is modeled by a microscopic Hamiltonian $\hat{H}$, given in the main text. Utilizing  $\hat{H}$, we get the following Heisenberg equations for the operators ($\hslash=1$) \cite{DharSen_BoundState_PhysRevB2006,DRoy_ElectronSelfConsistent_PhysRevB2007}:
\begin{gather}
    \frac{\partial \hat{b}_1(l)}{\partial t}=i\big[ \hat{b}_2(l)+\gamma_l\hat{c}_l \big], \label{S47}\\
    \frac{\partial \hat{b}_m(l)}{\partial t}=i\big[ \hat{b}_{m+1}(l)+\hat{b}_{m-1}(l) \big]~~~\text{for}~~m\neq 1,\label{S48}\\
    \frac{\partial \hat{c}_l}{\partial t}=i(\hat{c}_{l-1}+\hat{c}_{l+1})+i\gamma_l \hat{b}_1(l)~~~\text{for}~~l=1,2...,N,\label{S49}
\end{gather}
where we have taken $\gamma_o=1$. The open boundary condition on the wire imposes $\hat{c}_{0}=\hat{c}_{N+1}=0$. To integrate out the bath's degrees of freedom, we use retarded Green’s function operator $\hat{g}^{+}(l,t)=-i\theta(t)e^{-i\hat{H}^l_{B}t~}$ for each bath located at $l$.  The matrix elements of an operator $\hat{g}^{+}(l,t)$ in the real-space basis are given by
\begin{align}
   g^{+}_{m,n}(l,t)=-i\theta(t)\int_{0}^{\pi} dk \psi_k(m)\psi^{*}_k(n) e^{-i\omega_kt}.
\end{align}
Here, the energy-momentum dispersion $\omega_k=-2\cos{k}$ represents the single particle spectrum of $\hat{H}^{l}_B$.  The corresponding wavefunctions $\psi_k(m)$ in a semi-infinite TB chain are given by $\psi_k(m)=\sqrt{2/\pi}\sin{km}$, where $k\in(0,\pi)$.
These wavefunctions are used to define normal mode operators for the $l$th bath: $\hat{b}^\dagger_k(l)=\sum_m \psi_k(m) \hat{b}^\dagger_m(l)$, which diagonalizes  $\hat{H}^{l}_B$. Further, $\psi_k(m)$ satisfies the following two relations :  $\sum_{m\geq 1} \psi_k(m)\psi^*_{k'}(m)=\delta(k-k'),$ and $\int_{0}^{\pi} dk \psi_k(m)\psi^*_{k}(n)=\delta_{mn}$.
It is then easy to check that 
$\{\hat{b}_m(l),\hat{b}^\dagger_{n}(l')\}=\delta_{mn}\delta_{ll'} 
  $ gives $ \{\hat{b}_k(l),\hat{b}^\dagger_{k'}(l')\}=\delta(k-k')\delta_{ll'}$. Using Eqs.~\ref{S47}-\ref{S48}, we obtain the following equation of motion for $\hat{b}_{k}(l)$:
\begin{align}
    \frac{\partial \hat{b}_k(l)}{\partial t}=-i\omega_k\hat{b}_k(l)+i\gamma_l\psi^*_k(1)\hat{c}_l.
\end{align}
Integrating the above equation from an initial time $t_o$ to a later time $t$, we get,
\begin{align}
&\hat{b}_k(l,t)=\hat{b}_k(l,t_o)e^{i\omega_k(t_o-t)}+i\gamma_l\psi^*_k(1)\int_{t_o}^{t} d\tau \hat{c}_l(\tau)e^{-i\omega_k (t-\tau)},\nonumber\\
&\hat{b}_m(l,t)=i~\sum_{n}g_{m,n}^{+}(l,t-t_o) \hat{b}_{n}(l;t_o)\nonumber\\
&~~~~~~~~~~~~-~\gamma_l\int_{t_o}^{t} d\tau g_{m,1}^{+}(l,t-\tau)\hat{c}_l(\tau),~~\text{where}~t>t_o.\label{S52}
\end{align}
In the previous line, we have used the relation $\hat{b}^\dagger_m(l,t)=\int dk \psi^{*}_k(m) \hat{b}^\dagger_k(l,t)$.
Next, we substitute the expression for $\hat{b}_{1}(l,t)$ at a later time $t$ into Eq.~\ref{S49}. This will lead to the generalized quantum Langevin equations for the wire's degrees of freedom :
\begin{align}
\frac{\partial \hat{c}_l}{\partial t}=&i(\hat{c}_{l+1}+\hat{c}_{l-1})-i\hat{\eta}_l(t)\nonumber\\
&-i\int_{t_o}^{t} d\tau \Sigma_l^{+}(t-\tau)\hat{c}_l(\tau)~~~\text{for}~~l=1,2...,N\label{EQM_wire}
\end{align}
which is given in Eq.~\ref{QLE_wire}.
  We introduce the self-energy correction, $\Sigma_l^{+}(t)$ in the time domain and the noise operator, $\hat{\eta}_l(t)$ as:
\begin{align}
  &\hat{\eta}_l(t)=-i~\gamma_l\sum_{m}g_{1,m}^{+}(l,t-t_o) \hat{b}_{m}(l,t_o)~~\text{and},\nonumber\\
  &\Sigma_l^{+}(t)=~\gamma_l^2 g_{1,1}^{+}(l,t). \label{noise and self energy in the time domain}
\end{align}
They both appear because of a finite coupling of the wire with a bath located at site $l$. 
\subsection{Steady-state solutions and noise correlations }\label{Sec_2_1}
Next, we explore the charge transport through the wire in the steady state ($t \to \infty$). Let us assume that the wire was connected to the baths in the remote past ($t_o\to -\infty$). This simplifies the analysis by facilitating  Fourier transformation to the frequency domain.
The Fourier transformations from the time domain to the frequency domain for the operators and complex numbers are:
\begin{align}    &\hat{{c}}_l(\omega)=\frac{1}{2\pi}\int_{-\infty}^{\infty} dt e^{i\omega t} \hat{c}_l(t), ~ \hat{{\eta}}_l(\omega)=\frac{1}{2\pi}\int_{-\infty}^{\infty}  dt e^{i\omega t} \hat{\eta}_l(t),\nonumber\\
    &\text{and}~~ {\Sigma}_{l}^{+}(\omega)=~\gamma_l^2\int_{-\infty}^{\infty} dt e^{i\omega t}{ g}_{1,1}^{+}(l,t).\label{Fourier transformation relations}
\end{align}
Using the inverse Fourier transformation of the above relations, we transform the quantum Langevin equations in Eq.~\ref{EQM_wire} into
\begin{align}
    \hat{{c}}_{l-1}(\omega)+[\omega-{\Sigma}_l^{+}(\omega)]\hat{{c}}_{l}(\omega)+\hat{{c}}_{l+1}(\omega)=\hat{{\eta}}_l(\omega) \label{S56},
\end{align}
for $l=1,2,..., N$. The above equations are linear in the operators $\hat{{c}}_{l}(\omega)$ with the noises appearing as inhomogeneous terms on the right-hand side. Therefore, we can invert Eq.~\ref{S56} to get the steady-state solutions in the frequency domain as follows :
\begin{align}
    \hat{{c}}_{l}(\omega)=\sum_{l'=1}^{N} G^{+}_{ll'}(\omega)\hat{{\eta}}_{l'}(\omega)~~~\text{for}~~l=1,2...,N,\label{S57}
\end{align}
where $G^{+}_{ll'}(\omega)$ are the elements of ($\hat{Z}^{-1}$). The elements of the matrix $\hat{Z}$ have the following form : $ Z_{ll'}(\omega)= \delta_{l,l'-1}+[\omega-{\Sigma}_l^{+}(\omega)]\delta_{l,l'}+\delta_{l,l'+1}$.

Following the definition in Eq.~\ref{noise and self energy in the time domain}, the noise in the wire enters through the bath operators $\hat{b}_{m}(l,t_o)$. A bath coupled to the wire at site $l$ was initialized in a thermal state, which is characterized by the grand canonical ensemble with temperature $T$ and \textcolor{black}{chemical} potential $\mu_l$. Therefore, the normal mode operators $\hat{b}^\dagger_{k}(l),\hat{b}_{k}(l)$ satisfy the following average relations : $\langle \hat{b}^\dagger_{k}(l,t_o)\rangle=0$, $\langle \hat{b}_{k}(l,t_o)\rangle=0$ and 
$\langle \hat{b}^\dagger_{k}(l,t_o)\hat{b}_{k'}(l',t_o)\rangle=f_l(\omega_k)\delta(k-k')\delta_{l,l'}$.
Here, $f_l(\omega)$ represents the Fermi function, expressed as $f_l(\omega)=[e^{(\omega-\mu_l)/K_B T}+1]^{-1}$ and $K_B$ stands for the Boltzmann constant. As a consequence of the above relations, we immediately obtain $\langle\hat{\eta}_l(t)\rangle=\langle\hat{\eta}_l^\dagger(t)\rangle=0$. We are interested in finding the average charge currents across various bonds in the wire. This requires two-point correlations in noise operators. The noise correlations arising from a bath at site $l$, between any two times $t$ and $t'$  are given by
\begin{align}
     &\langle\hat{\eta}_l^\dagger(t) \hat{\eta}_{l}(t')\rangle \nonumber\\
     =&~\gamma_l^2\sum_{m,n}g_{1,m}^{+*}(l,t-t_o)g_{1,n}^{+}(l;t'-t_o) \langle \hat{b}_{m}^\dagger(l,t_o)\hat{b}_{n}(l,t_o)\rangle \nonumber\\
      =&~\gamma_l^2\sum_{m,n}g_{1,m}^{+*}(l,t-t_o)g_{1,n}^{+}(l;t'-t_o)  \int_{0}^{\pi}  dk \psi^*_k(m)\psi_{k}(n) f_l(\omega_k) \nonumber \\
 =&\gamma_l^2\int_{0}^{\pi}  dk  |\psi_k(1)|^2 f_l(\omega_k) e^{-i\omega_k(t'-t)}.
\end{align}
 Thus, the noises from the baths are colored. They arise due to the finite bandwidth and nonlinear dispersion of the structured TB baths.
In the frequency domain, the noise correlations get simplified to 
\begin{align}
    \langle\hat{\eta}_l^\dagger(\omega) \hat{\eta}_l(\omega')\rangle&=\frac{1}{(2\pi)^2}\int \int dt dt'e^{-i\omega t}e^{i\omega' t'}\langle\eta_l^\dagger(t) \eta_l(t')\rangle \nonumber\\
    &=\gamma_l^2 f_l(\omega) \rho_l(\omega)\delta(\omega-\omega'),\label{Noise_noise correlations in frequency domain}
\end{align}
where   $\rho_l(\omega_k)=|\psi_{k}(1)|^2/|\partial_k\omega_k|$ gives the density of states at the first site of the $l$th bath. For a semi-infinite TB chain, we get  $\rho_l(\omega_k)=\sqrt{4-\omega_k^2}/(2\pi)$.

We now calculate the self-energy correction in the frequency domain resulting from connecting a bath to the wire at site $l$. 
\begin{align}
    {\Sigma}_l^{+}(\omega)
    =& ~\gamma_l^2 \int_{-\infty}^{\infty} dt e^{i\omega t} g_{1,1}^{+}(l,t) \nonumber\\
     =& -i\gamma_l^2 \int_{-\infty}^{\infty} dt \theta(t)\,_{l}\langle 1|e^{i[\omega-\hat{H}^l_B]t}|1\rangle_{l}\nonumber\\
     =& -i\gamma_l^2 \int_{0}^{\infty} dt \lim_{\epsilon \to 0^{+}} \,_{l}\langle 1|e^{i[\omega-\hat{H}^l_B+i\epsilon]t}|1\rangle_{l} \nonumber \\
      =&\gamma_l^2 \lim_{\epsilon \to 0^{+}}\,_{l}\langle 1|[\omega-\hat{H}^l_B+i\epsilon]^{-1}|1\rangle_{l} \nonumber \\
   =& \frac{\gamma_l^2}{2}\big(\omega-i\sqrt{4-\omega^2}\big)=\Delta_l(\omega)-i\frac{\Gamma_l(\omega)}{2}.\label{self_energy}
\end{align}
where $\hat{b}^{\dagger}_{m}(l)|\varphi\rangle=|m\rangle_{l}$. In Eq.~\ref{self_energy}, we have separated the real and imaginary parts of the self-energy ${\Sigma}_l^{+}(\omega)$. The real part gives the frequency-dependent energy shift on the wire's onsite energies and is given by: $\Delta_l(\omega)=\gamma^2_l\omega/2$.
The imaginary part is always negative, and it introduces dissipation in the wire. Using the expression for $\rho_l(\omega)$, we can rewrite $\Gamma_l(\omega)=2\pi \gamma_l^2\rho_l(\omega)$.
We can relate the noise correlations with the dissipative component of the self-energy through the following (fluctuation-dissipation) relation:
\begin{align}
\langle\eta_l^\dagger(\omega) \eta_l(\omega')\rangle =\frac{\Gamma_l(\omega)}{2\pi} f_l(\omega) \delta(\omega-\omega').\label{fluctuation_dissipation relations}
\end{align}
\subsection{Charge currents in the steady state}\label{Sec_2_2}
 We again apply a continuity equation to obtain various currents in the system.
Let us define the local particle density at site $l$ in the wire as $n_l=\langle \hat{c}_l^\dagger \hat{c}_l \rangle$. We use the equations of motion in Eqs.~\ref{S47}-\ref{S49} to derive the time rate of change of local charge density at site $l$ :
\begin{align}
    \frac{\partial (en_l) }{\partial t}=&ie\gamma_l \langle \hat{c}_l^\dagger  \hat{b}_1(l)-\hat{b}_1^\dagger(l)\hat{c}_l\rangle +ie \langle \hat{c}_{l}^\dagger \hat{c}_{l-1}-\hat{c}_{l-1}^\dagger \hat{c}_{l} \rangle\nonumber\\
    &~~~-ie\langle \hat{c}_{l+1}^\dagger \hat{c}_l-\hat{c}_l^\dagger \hat{c}_{l+1} \rangle. \label{Continuity equation for particle}
\end{align}
Using the above expression, we identify two kinds of charge currents in the wire: 
\begin{gather}
    J_l=ie\langle c_{l+1}^\dagger c_l-c_l^\dagger c_{l+1} \rangle,\label{Eq63}\\
     I_l=ie\gamma_l\langle c_l^\dagger b_{1}(l)- b_{1}^\dagger(l) c_l \rangle=2e\gamma_l~\text{Im}\langle b_{1}^\dagger(l) c_l\rangle.\label{Eq64}
\end{gather}
The current flowing through the wire from left to right across the bond between the sites $l$ and $l+1$ is represented by $J_l$.
Similarly, the current entering the wire through the junction between the first site of the $l$th bath and site $l$ of the wire is denoted by $I_l$.
In the steady state, we utilize the noise correlations in the frequency domain to derive expressions for $J_l$ and $I_l$. From Eq.~\ref{Eq63}, we get :
\begin{align}
    J_l 
    =&ie\int \int d\omega d\omega' e^{i(\omega-\omega')t}\langle \hat{{c}}_{l+1}^\dagger(\omega) \hat{ {c}}_l(\omega')-\hat{{c}}_l^\dagger(\omega) \hat{{c}}_{l+1}(\omega') \rangle \nonumber \\
    =&ie\int d\omega  \sum_{l'} \big[ G_{l'l+1}^{-}(\omega)  G^{+}_{ll'}(\omega)- G_{l'l}^{-}(\omega)  G^{+}_{l+1l'}(\omega) \big]  \nonumber\\
    &~~~~~~~~~~~~~~~\times\gamma_{l'}^2\rho_{l'}(\omega) f_{l'}(\omega)\nonumber \\
     =&\frac{e}{2\pi}\int d\omega  \sum_{l'} \mathcal{F}_{ll'}(\omega) \big[f_{l}(\omega)-f_{l'}(\omega)\big]\label{particle current in the wire}. 
\end{align}
In the above equation, we  introduce $ G_{l'l}^{-}(\omega)\equiv G_{ll'}^{+*}(\omega)$ and 
$\mathcal{F}_{ll'}(\omega)=i[G_{l'l}^{-}(\omega)  G^{+}_{l+1l'}(\omega)- G_{l'l+1}^{-}(\omega)  G^{+}_{ll'}(\omega)]\Gamma_{l'}(\omega)$. We also use $\sum_{l'}\mathcal{F}_{ll'}(\omega)=0$, which can be demonstrated using the following two relations: 
\begin{align}
&i \sum_{l=1}^{N} G^{+}_{il}(\omega) \Gamma_l(\omega)G^{-}_{lj}(\omega)=G^{-}_{ij}(\omega)-G^{+}_{ij}(\omega),\\
& \text{Re}[G^{+}_{ll'}(\omega)]=\text{Re}[G^{-}_{ll'}(\omega)].
\end{align}
We then simplify the expression for the current $I_l$ in Eq.~\ref{Eq64} that flows in the wire from a bath located at site $l$.
\begin{align}
    I_l &=-2e\text{Im} \bigg[ \int \int d\omega d\omega' e^{i(\omega-\omega')t}   \bigg\langle \big\{\hat{{\eta}}^\dagger_l(\omega)\nonumber\\
    &~~~~~~~~~~~~~~~~~~~~~~~~~~~~~~~~~~+{\Sigma}_l^{+*}(\omega)\hat{{c}}^\dagger_{l}(\omega)\big\} \hat{ c}_l(\omega') \bigg\rangle \bigg] \nonumber
\end{align}
\begin{align}
& =-2e\text{Im} \bigg[ \int \int d\omega d\omega' e^{i(\omega-\omega')t}   \bigg\{ \sum_{j} G^{+}_{lj}(\omega)\langle \hat{{\eta}}^\dagger_l(\omega)\hat{{\eta}}_j(\omega') \rangle\nonumber\\
&~~~~~~~~~~~~~~~~~~+{\Sigma}_l^{+*}(\omega)  \sum_{i,j} G^{+*}_{li}(\omega)G^{+}_{lj}(\omega')\langle \hat{{\eta}}^\dagger_i(\omega)\hat{{\eta}}_j(\omega') \rangle \bigg\} \bigg] \nonumber \\
  & =-2e \int d\omega  \bigg\{ ~\text{Im}\big[G^{+}_{ll}(\omega)\big] \gamma_l^2\rho_l(\omega)f_l(\omega)\nonumber\\
 &~~~~~~~~~~~~~~~~~~~~~~~~+\text{Im}\big[{\Sigma}_l^{+*}(\omega)\big]  \sum_{l'} |G^{+}_{ll'}(\omega)|^2 \gamma_{l'}^2\rho_{l'}(\omega)f_{l'}(\omega)\bigg\}  \nonumber\\
  & =-\frac{e}{2\pi} \int d\omega  \bigg\{-\sum_{l'}G^{+}_{ll'}(\omega)\Gamma_{l'}(\omega)G^{-}_{l'l}(\omega) f_l(\omega)\Gamma_l(\omega)\nonumber\\
  &~~~~~~~~~~~~~~~~~~~~~~~~~~~~+\Gamma_l(\omega) \sum_{l'} |G^{+}_{ll'}(\omega)|^2 f_{l'}(\omega)\Gamma_{l'}(\omega)\bigg\}  \nonumber\\
&=\frac{e}{2\pi} \sum_{l'}\int d\omega  \mathcal{T}_{ll'} (\omega)[f_l(\omega)-f_{l'}(\omega)] .\label{particle_current_wire_to_bath}
\end{align}
Here, we define the transmission coefficients between two baths located at sites $l$ and $l'$ by $\mathcal{T}_{ll'} (\omega)=|G^{+}_{ll'}(\omega)|^2 \Gamma_{l}(\omega)\Gamma_{l'}(\omega)$.

The current formulae in Eq.~\ref{particle current in the wire} and Eq.~\ref{particle_current_wire_to_bath} are expressed as integrals over all frequencies in the Fourier domain. For analytical purposes, we can further simplify these expressions. We give a detailed derivation for $I_l$  expanded in all powers of $\delta\mu_l=\mu_l-~\bar\mu$. 
Let us introduce the integration variable $z=\beta(\omega-\bar\mu)$, where $\beta$ is reciprocal of the temperature $T$, i.e., $ \beta=(K_B T)^{-1}$. We perform a Taylor series expansion of $f_l(\omega)$ about $\mu_l=\bar\mu$. Then, the differences in the Fermi functions can be written as follows:
\begin{align}
    &f_{l}(\omega)-f_{l'}(\omega) = \sum_{n\geq 1}^{\infty} \frac{1}{n!}\frac{\partial^n f_l(\omega)}{\partial \mu_l^n}\bigg|_{\mu_l=\bar\mu} (\delta \mu_l^n-\delta \mu_{l'}^n)\nonumber\\
    &=\sum_{n\geq 1}^{\infty} \frac{(-1)^n \beta^n}{n!}\frac{\partial^n}{\partial z^n} (1+e^z)^{-1} (\delta \mu_{l}^n-\delta \mu_{l'}^n).
\end{align}
Similarly, we carry out a series expansion for the transmission coefficients $\mathcal{T}_{ll'}(\omega)$ about $~\omega=\bar\mu$ and write Eq.~\ref{particle_current_wire_to_bath} as:
\begin{align}
     I_{l}
    =&\frac{e}{2\pi}\sum_{l'=1}^{N}\sum_{n\geq 1}^{\infty}\sum_{m=0}^{\infty} \frac{(-1)^n}{n! m!} \beta^{n-(m+1)}\mathcal{T}_{ll',m}  \nonumber\\
    & ~~~~~~~~~~~  \times(\delta \mu_l^n-\delta \mu_{l'}^n) \int_{-\infty}^{\infty} dz~ z^m\frac{\partial^n }{\partial z^n}(1+e^z)^{-1}. \label{A3}
\end{align}
Here, we use the notation $\mathcal{T}_{ll',m}=\frac{\partial^{m}}{\partial{\omega^m}}\mathcal{T}_{ll'}(\omega)\big|_{\omega=\bar\mu~}$. We choose a temperature $T \ll \bar\mu/K_B $,  such that the terms in the series containing $\beta^{n-(m+1)}$ for $n< m+1$ can be neglected. Now, for $n \geq m+1$, the above integration (by repetitive integrations by-parts) gives:
\begin{align}
  \int_{-\infty}^{\infty} dz~  z^m\frac{\partial^n }{\partial z^n}(1+e^z)^{-1}= \delta_{m,n-1} (-1)^n (n-1)!,
\end{align}
where $n!$ denotes the factorial value of an integer $n$. Therefore, at temperatures $T \ll \bar\mu/K_B $, the current $I_l$  due to the applied \textcolor{black}{chemical} potential biases $\delta\mu_l$ can be written as 
\begin{align}
   I_{l}\approx \frac{e}{2\pi}\sum_{l'=1}^{N}\sum_{n\geq 1}^{\infty} \frac{1}{~ n!} \mathcal{T}_{ll',n-1}(\delta \mu_l^n-\delta \mu_{l'}^n).\label{S72}
\end{align}
Similarly, we can get 
\begin{align}
       J_{l}\approx \frac{e}{2\pi}\sum_{l'=1}^{N}\sum_{n\geq 1}^{\infty} \frac{1}{~ n!} \mathcal{F}_{ll',n-1}(\delta \mu_l^n-\delta \mu_{l'}^n),\label{S73}
\end{align}
where $\mathcal{F}_{ll',m}=\frac{\partial^{m}}{\partial{\omega^m}}\mathcal{F}_{ll'}(\omega)\big|_{\omega=\bar\mu~}$. We have used $\mathcal{T}_{ll',0}\equiv \mathcal{T}_{ll'}$, and $\mathcal{F}_{ll',0}\equiv \mathcal{F}_{ll'}$ in the main text. In the linear response regime, we truncate the series in Eqs.~\ref{S72}-\ref{S73}  after $n=1$.
\subsection{ One S-D dimer: linear and nonlinear response regime}\label{Sec_2_3}
We consider $N=4$, and a single S-D dimer ($N_i/2=1$).
The transmission coefficients $\mathcal{T}_{ll'}$ are symmetric in  $l$ and $l'$, and we derive them fully analytically as follows: 
\begin{gather}
\mathcal{T}_{24}=\mathcal{T}_{31}=\frac{4\gamma^2}{\mathcal{D}(\bar\mu)}(1-\bar\mu^2/4),\\
\mathcal{T}_{23}=\frac{4\gamma^4}{\mathcal{D}(\bar\mu)}(1-\bar\mu^2/4),\mathcal{T}_{14}=\frac{4}{\mathcal{D}(\bar\mu)}(1-\bar\mu^2/4), \\
\mathcal{T}_{21}=\mathcal{T}_{34}=\frac{4\gamma^2}{\mathcal{D}(\bar\mu)}(1-\bar\mu^2/4)[1+\gamma^4-\gamma^2(\bar\mu^2-2)],
\end{gather}
where $\mathcal{D}(\bar\mu)=(2+2\gamma^2+\gamma^4)^2-(1+2\gamma^2+5\gamma^4+2\gamma^6)\bar\mu^2+\gamma^4\bar\mu^4$. \textcolor{black}{Using the S and D current constraints, i.e., $I_l=(-1)^l I_g$ for $l=2,3$ in the linear response regime, we  obtain two linear equations}. They give the solutions : $\mu_2=\bar\mu+\Delta_\mu$ and $\mu_3=\bar\mu-\Delta_\mu$, where
\begin{align}
     \Delta_\mu &=\frac{2\pi (I_g/e)+(\mathcal{T}_{34
     }-\mathcal{T}_{13})(\mu_{\text{L}}-\bar\mu)}{\mathcal{T}_{13}+\mathcal{T}_{34}+2\mathcal{T}_{23}}\nonumber\\
     &=\frac{\pi \mathcal{D}(\bar\mu)I_g/e+\gamma^4(1-\bar\mu^2/4)(\gamma^2+2-\bar\mu^2)\delta\mu}{2\gamma^2(1-\bar\mu^2/4)(2+\gamma^4+4\gamma^2-\gamma^2\bar\mu^2)}, \label{S77}
\end{align}
\begin{figure}[h!]
\centering
\includegraphics[width=0.47\textwidth]{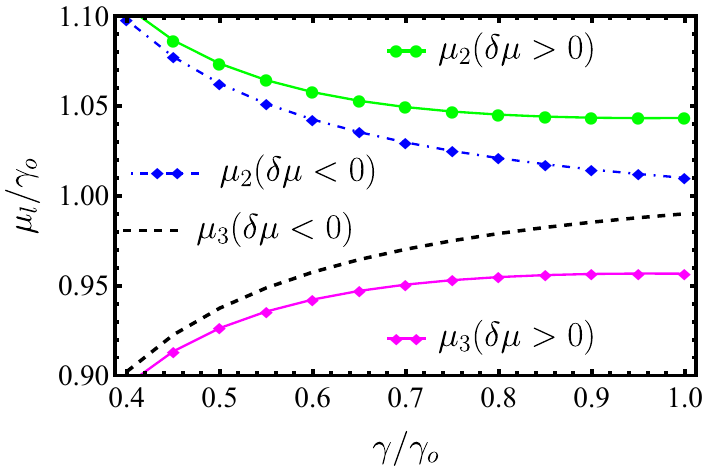}
\caption{ \textcolor{black}{Variation of \textcolor{black}{chemical} potentials $\mu_2$ and $\mu_3$ with $\gamma$ in the linear
response regime of a short quantum wire (N = 4), where $\bar\mu/\gamma_o=1$, $I_g/I_o=0.3$, $\delta\mu/\gamma_o=0.1$ and $e=1$.}}\label{variation with gamma}
\end{figure}
\textcolor{black}{The above expression leads to Eq.\ref{Delta_muEq} once we expand it in powers of $\gamma$. Further, a variation of $\mu_2,\mu_3$ with $\gamma$ is shown in Fig.~\ref{variation with gamma}.} The current flowing in the wire from the left end is 
$I_{1}=G_4(\delta\mu/e)-\delta I$, where
\begin{align}
G_4&=\frac{2\mathcal{T}_{34}\mathcal{T}_{13}+2\mathcal{T}_{23}\mathcal{T}_{14}+(\mathcal{T}_{23}+\mathcal{T}_{14})(\mathcal{T}_{34}+\mathcal{T}_{13})}{2\pi (\mathcal{T}_{13}+\mathcal{T}_{3
4}+2\mathcal{T}_{23})}\nonumber\\
&=e^2\frac{(4-\bar\mu^2)(2+\gamma^4+2\gamma^2-\gamma^2\bar\mu^2)(\gamma^2+1)^2}{2\pi\mathcal{D}(\bar\mu)(2+\gamma^4+4\gamma^2-\gamma^2\bar\mu^2)},\\
 \delta I &=\frac{(\mathcal{T}_{12}-\mathcal{T}_{13})I_g}{ (\mathcal{T}_{13}+\mathcal{T}_{3
4}+2\mathcal{T}_{23})}=\frac{(\gamma^4+2\gamma^2-\gamma^2\bar\mu^2)}{(2+\gamma^4+4\gamma^2-\gamma^2\bar\mu^2)}I_g.
\end{align}
The above equations give Eq.\ref{G4_Eq} in the main text. Equating $\delta I=I_g/3$, we get only one real solution for $\gamma$, which is given by
\begin{align}
    \gamma_c^2=\frac{\bar\mu^2-1+\sqrt{(\bar\mu^2-1)^2+4}}{2},
\end{align}
e.g., for $\bar\mu=1$, we get $\gamma_c=1$, which is in the non-perturbative regime of the wire-bath coupling.
Next, we consider the nonlinear current response. In the series expansion of $I_l$ in Eq.~\ref{S72}, we retain terms up to the third order ($n=3$) in  $\delta\mu_l$. This gives  the following pair of  nonlinear S-D equations with  $\delta\mu_2$ and $\delta\mu_3$ as the unknown variables:
\begin{align}
   &\frac{v''}{6}\delta\mu_2^3-\frac{w''}{6}\delta\mu_3^3+ \frac{v'}{2}\delta\mu_2^2-\frac{w'}{2}\delta\mu_3^2+v\delta\mu_2-w\delta\mu_3=\xi_2,\nonumber\\
   & \frac{v''}{6}\delta\mu_3^3-\frac{w''}{6}\delta\mu_2^3+ \frac{v'}{2}\delta\mu_3^2-\frac{w'}{2}\delta\mu_2^2+v\delta\mu_3-w\delta\mu_2=\xi_3,\label{S81} \\
   &\xi_l =(-1)^l  \frac{2\pi I_g}{e}+\frac{\delta\mu}{2}(\mathcal{T}_{l1}-\mathcal{T}_{l4})+\frac{\delta\mu^2}{8}(\mathcal{T}_{l1,1}+\mathcal{T}_{l4,1})\nonumber\\
   &~~~~~~~~~~~~+\frac{\delta\mu^3}{48}(\mathcal{T}_{l1,2}-\mathcal{T}_{l4,2}),
\end{align}
where $v\equiv v(\bar\mu)=(\mathcal{T}_{31}+\mathcal{T}_{32}+\mathcal{T}_{34})$,  $w\equiv w(\bar\mu)=\mathcal{T}_{32}$; $v'=\partial_{\bar\mu}v$,$v''=\frac{\partial^2 v}{\partial \bar\mu^2}$  and similarly for $w$. Due to the coupled nonlinearity for $\delta\mu_2$ and $\delta\mu_3$ in Eq.~\ref{S81}, we can get many pairs of complex solutions. Out of these, only one real pair $(\mu_2^{(3)},\mu_3^{(3)})$ is of physical relevance which coincides with the linear response results in Eq.~\ref{S77} for $\delta\mu_2^{(3)},\delta\mu_3^{(3)} \leq \delta\mu$ (Fig.~\ref{nonlinear_fig}). We use this physical pair to compute the magnitude of the incoming and outgoing current $|I_1|$ in the forward and reverse bias. We show the corresponding relative nonreciprocity in Fig.~\ref{Fig1}(c). 
\begin{figure}[h!]
\centering
\includegraphics[width=0.47\textwidth]{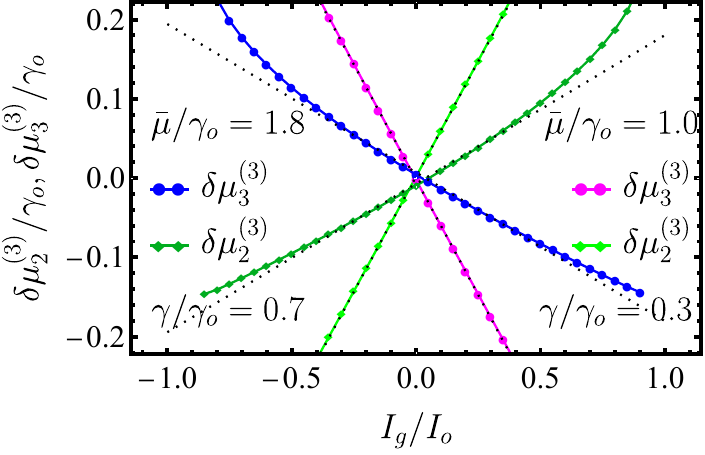}
\caption{Numerical roots $(\mu_2^{(3)},\mu_3^{(3)})$ of the nonlinear Eq.~\ref{S81}. The dotted black lines show the corresponding solutions obtained within the linear response analysis. In the plot, we fix $\delta\mu/\gamma_o=0.1$ and $e=1$.}\label{nonlinear_fig}
\end{figure}
\subsection{Linear response regime for the multiple dimers}\label{Sec_2_4}
\begin{figure*}
\includegraphics[height=4.6cm,width=17.5cm]{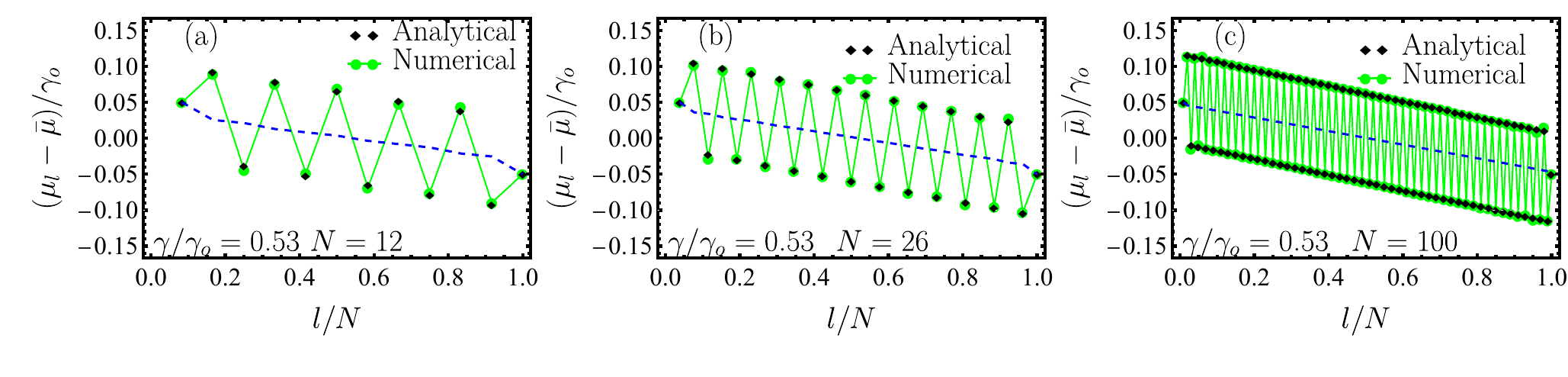}
\caption{\textcolor{black}{Chemical} potential profile in a finite quantum wire with increasing size $N$: (a) N=12, (b) N=26, and (c) N=100. The blue dashed line shows the corresponding plot for $I_g=0$, with $\gamma\neq0$.  Here, $I_g/I_o=0.3$, $\bar\mu/\gamma_o=1.4$, $\delta\mu/\gamma_o=0.1$, and $e=1$.}
\label{Fig8}
\end{figure*}
Next, we focus on the multiple S-D baths in a longer quantum wire ($N\gg 4$).
We restrict our calculations to the linear response regime for analytical traceability and consider a large number of S-D dimers with $N_i =N-2$. Let us further assume  $\gamma<1$, which gives $l_c > 1$. The following parameters $\sigma,l_c,\alpha (=\alpha_R+i\alpha_I)$ are introduced in the main text. We show the following \textcolor{black}{chemical} potentials  (Fig.~\ref{Fig8}) of the middle baths fix the source and drain currents self-consistently at the bulk sites, which are located far away from the boundaries:
\begin{align}
    &\mu_l=\mu_{\text{L}}-\phi+\Big[\frac{l_c}{2\sigma}+(-1)^l\nabla\Big]\frac{eI_g}{2} -\frac{2\phi}{l_c} (l-2), \label{Achempot__multipleSD} \\
    &\phi=\frac{\delta\mu+eI_g l_c/(2\sigma)}{2[1+(N-3)/l_c]},~~\text{and}~~ \nabla= \frac{\coth^2{\alpha_R}}{2 \sigma},
\end{align}
for $l=2,3,...,N-1$.  First, we compute  $I_l$ at an odd site $l$ in the bulk, such that $1 \ll l\ll N$. Within the linear response regime, we get
\begin{align}
    I_l 
    &=\frac{e}{2\pi}\sum_{l'\neq 1,N}\mathcal{T}_{ll'}(\delta \mu_l-\delta \mu_{l'})+\frac{e}{2\pi}\mathcal{T}_{l1}(\delta \mu_l-\delta \mu_{L})\nonumber\\
    &~~~~~~~~~~~~~~+\frac{e}{2\pi}\mathcal{T}_{lN}(\delta \mu_l-\delta \mu_{R})\nonumber\\
    &\approx\frac{e}{2\pi}\sum_{l'\neq 1,N}\mathcal{T}_{ll'}\Big[-\frac{eI_g}{2}\nabla-(-1)^{l'}\frac{eI_g}{2}\nabla -\frac{2\phi}{l_c} (l-l')\Big]\nonumber\\
    &=-\frac{e^2I_g}{2\pi}\nabla\sum_{l'=2,4,...}^{N-2}\mathcal{T}_{ll'}-\frac{2\phi}{l_c} \frac{e}{2\pi}\sum_{l'\neq 1,N}\mathcal{T}_{ll'} (l-l').
\end{align}
The boundary terms in the above calculations are neglected in comparison to the other terms since the corresponding transmission coefficients are negligibly small for $1 \ll l\ll N$. Again, for all the points $l,l'$ within the bulk of the wire,
we get \cite{DRoy_ElectronSelfConsistent_PhysRevB2007}: 
\begin{align}
\mathcal{T}_{ll'}=\frac{\pi^2\gamma^4\rho_l(\bar\mu)\rho_{l'}(\bar\mu)}{|\sinh{\alpha}|^2}e^{-2\alpha_R|l-l'|},
\end{align}
in the limit $N\to \infty$. In this limit, we can ideally get:
\begin{align}
   & \lim_{(N-l)\to \infty}\lim_{l\to \infty}\sum_{l'\neq 1,N}e^{-2\alpha_R|l-l'|}(l-l')=0,\label{S86}\\
    & \lim_{(N-l)\to \infty}\lim_{l\to \infty}\sum_{l'=2,4,...}^{N-2}e^{-2\alpha_R|l-l'|}=\frac{1}{\sinh{2\alpha_R}}\label{S87}.
\end{align}
Therefore, at an odd site $l$ in the bulk of the wire, we get 
\begin{align}
    I_l 
    &=-\frac{e^2I_g}{2\pi}\frac{\pi^2\gamma^4\rho^2_l(\bar\mu)}{|\sinh{\alpha}|^2}\frac{\nabla}{\sinh{2\alpha_R}}=-I_g.
\end{align}
We have used $\sigma=e^2\sin^2{\alpha_I}\coth{\alpha_R}/(2\pi |\sinh{\alpha}|^2)$, and $\rho_l(\bar\mu)=2\sin{\alpha_I\sinh{\alpha_R}}/(\pi\gamma^2)$. Similarly, we can get $I_l=I_g$ for all even points $l$ in the bulk.

Let us calculate the inter-bond current, $J_{\text{D}}=J_l$ ($l$ is odd) within the wire. In the linear response regime,
\begin{align}
    J_{\text{D}} 
     =&\frac{e}{2\pi}\sum_{l'\neq 1,N}\mathcal{F}_{ll'}(\delta \mu_l-\delta \mu_{l'})+\frac{e}{2\pi}\mathcal{F}_{l1}(\delta \mu_l-\delta \mu_{L})\nonumber\\
    &~~~~~~~~~~~~~~+\frac{e}{2\pi}\mathcal{F}_{lN}(\delta \mu_l-\delta \mu_{R})\nonumber\\
     \approx&-\frac{e^2I_g}{2\pi}\nabla\sum_{l'=2,4,...}^{N-2}\mathcal{F}_{ll'}-\frac{2\phi}{l_c} \frac{e}{2\pi}\sum_{l'\neq 1,N}\mathcal{F}_{ll'} (l-l').\label{S90}
\end{align}
 Now, for all points $l,l'$ within the bulk of the wire, we can get \cite{DRoy_ElectronSelfConsistent_PhysRevB2007}:
\begin{align}
&\mathcal{F}_{ll'}=-\frac{\pi\gamma^2\rho_{l}(\bar\mu)}{|\sinh{\alpha}|^2}e^{-(|l+1-l'|+|l-l'|)\alpha_R} \nonumber\\
&~~~~~~~~~~~~~~~~~~~~~~~~~~~~~\times\sin{\big(|l+1-l'|-|l-l'|\big)\alpha_I},
\end{align}  
for $N\to \infty$. Using the above relation, we evaluate the following two summations :
\begin{align}
   & \lim_{(N-l)\to \infty}\lim_{l\to \infty}\sum_{l'\neq 1,N}\mathcal{F}_{ll'}(l-l')\nonumber\\
   &=-\frac{\pi\gamma^2\rho_{l}(\bar\mu)}{|\sinh{\alpha}|^2}\frac{\cosh{\alpha_R}\sin\alpha_I}{2(\sinh{\alpha_R})^2}=-2\pi \frac{\sigma}{e^2},\\
    &\lim_{(N-l)\to \infty}\lim_{l\to \infty}\sum_{l'=2,4,...}^{N-2}\mathcal{F}_{ll'}=\frac{\pi\gamma^2\rho_{l}(\bar\mu)}{|\sinh{\alpha}|^2}\frac{\sin\alpha_I}{2\cosh{\alpha_R}}.
\end{align}
Substituting the above expressions into Eq.~\ref{S90}, we obtain an expression for $J_{\text{D}}$ as follows: 
\begin{align}
    J_{\text{D}} 
     =&-\frac{e^2I_g}{2\pi}\nabla\frac{\pi\gamma^2\rho_{l}(\bar\mu)}{|\sinh{\alpha}|^2}\frac{\sin\alpha_I}{2\cosh{\alpha_R}}+\frac{2\phi \sigma}{el_c} \nonumber\\
     =&-\frac{I_g}{2}+\frac{\sigma(\delta\mu/e)+I_g l_c/2}{l_c+(N-3)}.
\end{align}
The charge continuity equation in the steady state further gives $J_{\text{D}}=I_{1}=-I_{N}$. The above equation leads to Eq.~\ref{multiple SD current}. For the intra-bond currents, we choose $l$ at even sites on the wire, i.e., $J_{\text{S}}=J_l$ (even $l$). This yields :
\begin{align}
    J_{\text{S}} 
     =\frac{I_g}{2}+\frac{\sigma(\delta\mu/e)+I_g l_c/2}{l_c+(N-3)},
\end{align}
which is similar to Eq.~\ref{intrabond_lattice} in App.~\ref{Sec_1_3}.
Next, we inspect the local thermal equilibrium at the middle sites of the wire. For that, we evaluate the deviation, $\delta n_l$, of the local particle density (at any point $l$) from its corresponding equilibrium density. The local equilibrium density is obtained by imposing the condition: $\mu_{l'}=\mu_l$ for all $l'\in\{1,2,..., N\}$. We obtain the following expression for  $\delta n_l$ in the linear response regime:
\begin{align}
    \delta n_l&=\gamma^2\rho_l(\omega)\sum_{l'=1}^{N}\int d\omega |G^{+}_{ll'}(\omega)|^2    [f_{l'}(\omega)-f_l(\omega)]\nonumber\\
    &=\gamma^2\rho_l(\bar\mu)\sum_{l'} |G^{+}_{ll'}(\bar\mu)|^2 (\mu_{l'}-\mu_l).
\end{align}
For all points within the bulk of the wire, $G^{+}_{ll'}(\bar\mu)= (-1)^{|l-l'|} e^{-|l-l'|\alpha}/(2\sinh{\alpha}) $ for $N\to \infty$. Then, evaluating the above summation, we obtain
\begin{align}
    \delta n_l= (-1)^{l+1}\frac{ I_g}{2e\pi\gamma^2 \rho_l(\bar\mu)}.
\end{align}
We notice that $\delta n_l \neq 0$ as long as $I_g \neq 0$, i.e., the current driven nonequilibrium steady state does not provide local thermal equilibrium within the bulk of the wire.
 
{\it{Heat dissipation:}} The driven wire dissipates its energy in the form of heat into the baths (Joule heating). We see how the presence of driving currents $I_g$ affects the heat dissipation. 
We  particularly look into the heat energy lost per site within the bulk of the wire. For that, we need to calculate the heat
current flowing out of the wire from a site $l$ into the bath coupled at that site. It is defined as follows: $h_l=u_l-\mu_l (-I_l)$, where $u_l=i\gamma_l\langle (c^\dagger_{l+1}+c^\dagger_{l-1})b_1(l)-b^\dagger_1(l)(c_{l+1}+c_{l-1})\rangle$ denotes the corresponding flow of the energy current \cite{DRoy_ElectronSelfConsistent_PhysRevB2007}.
Like before, we consider a temperature $T\ll \bar\mu/K_B$ and retain the first non-zero terms in the series expansion of $h_{l}$ about $\mu_l=\bar\mu$. In this procedure, we get :
\begin{align}
    h_{l}=&\frac{1}{4\pi}\sum_{l'}\mathcal{T}_{ll'}(\mu_{l'}-\mu_{l})^2=\bigg[\sigma\bigg(\frac{2\phi}{el_c}\bigg)^2+\nabla \frac{I_g^2}{2}\bigg].\label{heat_current_multiple dimers}
\end{align} 
In the above expression, we use Eqs.~\ref{S86}-\ref{S87} and the following  relation:
\begin{align}
   & \lim_{(N-l)\to \infty}\lim_{l\to \infty}\sum_{l'\neq 1,N}e^{-2\alpha_R|l-l'|}(l-l')^2=\frac{\cosh\alpha_R}{2(\sinh\alpha_R)^3}.\label{S98}
\end{align}
We find the right hand side of Eq.~\ref{heat_current_multiple dimers}  is equal to $[J_{\text{S}}(\mu_{l-1}-\mu_{l})+J_{\text{D}}(\mu_{l}-\mu_{l+1})]/2$.  
The first term in Eq.\ref{heat_current_multiple dimers}  falls as $1/N^2$ (for $N\gg l_c$) as we increase the size of the wire (the diffusive component). 
We further notice that the nonreciprocal charge current affects only the first term in $h_l$. The second term in $h_l$  is $N$-independent. For a fixed $\delta\mu$, such a term in $h_l$ causes the overall heat dissipation in the wire to grow proportional to $N$. This is not surprising since the uniform driving along the wire will eventually cause more heating. For smaller values of $\gamma \ll 1$, we get 
\begin{align}
h_{l}\approx \frac{1}{\sigma}\bigg[\bigg(\frac{2\phi \sigma}{el_c}\bigg)^2+l_c^2\frac{I_g^2}{4}\bigg],
\end{align} 
where $\sigma^{-1}\approx 2\pi/(e^2l_c)$, and $l_c\approx 2/\gamma^2$.

\subsection{Role of contact scattering on  current nonreciprocity}\label{Contact_scattering_appendix}
Previously, we have chosen $\gamma_1/\gamma_o=\gamma_N/\gamma_o= 1$ for the analytical simplicity within the linear response regime. In this subsection, we  consider imperfect contacts between the wire and the L, R-baths, i.e.,  $\gamma_1/\gamma_o=\gamma_N/\gamma_o\neq 1$. This induces contact scattering at the boundaries, which influences the nonreciprocal current transport through the wire. 
\begin{figure}[h!]
\centering
\includegraphics[width=0.47\textwidth]{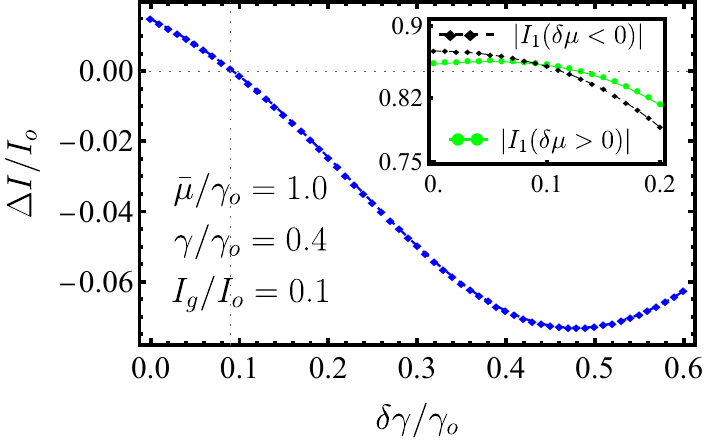}
\caption{The variation of the current nonreciprocity with $\delta \gamma$ in a single S-D dimer. We fix $\delta\mu/\gamma_o=0.1$ and $e=1$.}\label{conatct_1SD_nonreciprocity_fig}
\end{figure}

\begin{figure}[h!]
\centering
\includegraphics[width=0.47\textwidth]{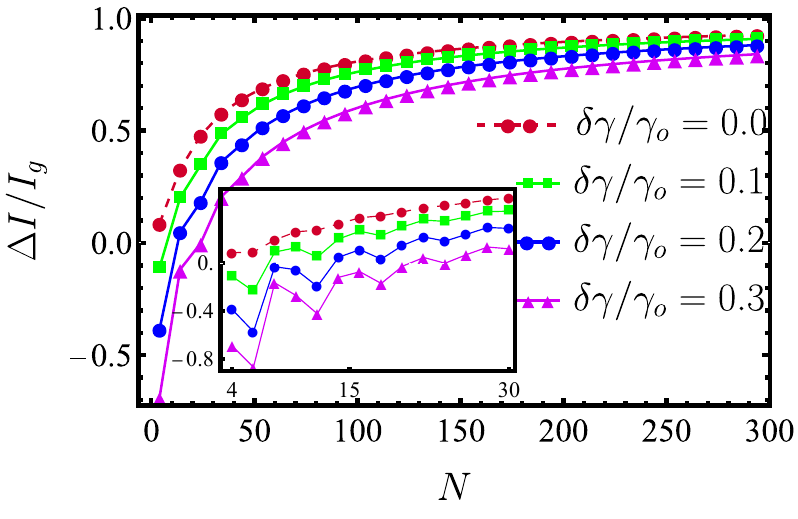}
\caption{Nonreciprocity in the charge current vs $N$ for different values of $\delta\gamma$ with  multiple S-D dimers. Parameters of the plot are $\bar\mu/\gamma_o=1.0$, $\delta\mu/\gamma_o=0.1$, $\gamma/\gamma_o=0.3$, $I_g/I_o=0.05$, and $e=1$. }\label{conatct_Multiple_SD_nonreciprocity_fig}
\end{figure}

We first take up a single S-D dimer ($N_i/2=1$).
In Fig.\ref{conatct_1SD_nonreciprocity_fig}, we have shown that the nonreciprocity $\Delta I=|I_1(\delta\mu<0)|-|I_1(\delta\mu>0)|$ value gradually decreases to zero with increasing $\delta\gamma$, where $\delta\gamma=\gamma_1-\gamma_o=\gamma_N-\gamma_o$.
 Beyond this point, if we further increase $\delta\gamma$, the current flowing in the forward bias $I_1(\delta\mu>0)$ becomes more than that of flowing in the reverse bias (inset of the Fig.~\ref{conatct_1SD_nonreciprocity_fig}). Next, we take $N_i/2=N/2-1$ pairs of S-D dimers in the wire. In Fig.\ref{conatct_Multiple_SD_nonreciprocity_fig}, we display the variation in $\Delta I$ with increasing $N$ for different values of $\delta \gamma$. For a sufficiently long wire, i.e.,  $N \gg l_c$, we notice $\Delta I$ asymptotically approaches $I_g$, irrespective of the value of $\delta \gamma$. In the intermediate length scales ($N \gtrsim l_c$), $\Delta I$ is more for a given $N$ with smaller $\delta\gamma$. Based on this observation, we infer that the introduction of contact scattering effectively enhances the scattering length scale $l_c$ in the quantum wire.   
The interplay of quantum coherence becomes more prominent for fewer S-D baths $N_i\sim 2$. We notice the oscillatory rise of  $\Delta I$ from negative values in smaller wire sizes, which can further lead to reciprocal charge transport ($\Delta I=0$) for a given $N\gtrsim 4$.

\section{Interplay between quantum coherence and nonreciprocal charge transport}\label{Sec_3}

In the previous section of the appendix, we observe how contact scattering can significantly affect the nonreciprocal current behavior in a short quantum wire.
In this section, we study the special limit of a single S-D dimer in a long wire, which can give reciprocal charge transport as discussed in Sec.~\ref{Sec_QCoh}. We couple one S and D bath to the wire at sites $l=2, N-1$, respectively. Therefore, we have $\gamma_l=(\delta_{l,2}+\delta_{l,N-1})\gamma$ for $l=2,3,...,N-1$. Once again, we fix the couplings at end contacts by $\gamma_1/\gamma_o=\gamma_N/\gamma=1$. In App.~\ref{Appendix_One S-D in a long ballistic quantum wire}, we derive analytical expressions for the \textcolor{black}{chemical} potentials of the S-D baths and the charge current $I_1$ in the wire. In App.~\ref{Appendix_One S-D in a long ballistic classical wire}, we get the nonreciprocal charge transport in a corresponding classical transport channel within the master equation description.
\subsection{Quantum modeling in linear response regime}\label{Appendix_One S-D in a long ballistic quantum wire}
The S-D currents give the following two equations, which fix $\mu_2$ and $\mu_{N-1}$ self-consistently:
 
\begin{align}
   & 2\pi (I_g/e)= (\mathcal{T}_{21}+\mathcal{T}_{2N-1}+\mathcal{T}_{2N})\delta\mu_2\nonumber\\
   &~~~~~~~~~~~~~~~~~~~~~~~~-\mathcal{T}_{2N-1}\delta\mu_{N-1}-(\mathcal{T}_{21}-\mathcal{T}_{2N})\delta\mu_{\text{L}},\label{EQ1_SD_in_long_ballistic} \\
    &-2\pi (I_g/e)= (\mathcal{T}_{1N-1}+\mathcal{T}_{2N-1}+\mathcal{T}_{NN-1})\delta\mu_{N-1}\nonumber\\
    &~~~~~~~~~~~~~~~~~~-\mathcal{T}_{2N-1}\delta\mu_{2}-(\mathcal{T}_{1N-1}-\mathcal{T}_{NN-1})\delta\mu_{\text{L}}.\label{EQ2_SD_in_long_ballistic}
\end{align}
We use the following relations: $\mathcal{T}_{ll'}=\mathcal{T}_{l'l}$, $\mathcal{T}_{12}=\mathcal{T}_{NN-1}$ and $\mathcal{T}_{1N-1}=\mathcal{T}_{2N}$, which  immediately give $\delta\mu_2=-\delta\mu_{N-1}=\Delta_{\mu,N}$. From Eq.~\ref{EQ1_SD_in_long_ballistic}, we obtain:
\begin{align}
    \Delta_{\mu,N} = \frac{2\pi (I_g/e)+(\mathcal{T}_{21}-\mathcal{T}_{2N})\delta\mu/2}{\mathcal{T}_{21}+2\mathcal{T}_{2N-1}+\mathcal{T}_{2N}}.
\end{align}
The current flowing in the wire can be written as:
\begin{align}
    I_1=e\frac{ (\mathcal{T}_{12}+\mathcal{T}_{1N-1}+2\mathcal{T}_{1N})\delta\mu_{\text{L}}-(\mathcal{T}_{12}-\mathcal{T}_{1N-1}) \Delta_{\mu,N}}{2\pi} ,\nonumber
\end{align}
which yields : $I_1=\tilde{G}_{N}(\delta\mu/e)-\Delta j/2$, where
\begin{align}
   &\tilde{G}_{N}=e^2\frac{\splitfrac{(\mathcal{T}_{12}+2\mathcal{T}_{1N}+\mathcal{T}_{1N-1})(\mathcal{T}_{21}+2\mathcal{T}_{2N-1}}{+\mathcal{T}_{2N})-(\mathcal{T}_{12}-\mathcal{T}_{1N-1})(\mathcal{T}_{21}-\mathcal{T}_{2N})}}{4\pi(\mathcal{T}_{21}+2\mathcal{T}_{2N-1}+\mathcal{T}_{2N})},\nonumber\\
    & \Delta j=2\frac{\mathcal{T}_{12}-\mathcal{T}_{1N-1}}{\mathcal{T}_{21}+2\mathcal{T}_{2N-1}+\mathcal{T}_{2N}}I_g.
\end{align}
We only need to evaluate the following four transmission coefficients: $\mathcal{T}_{21}, \mathcal{T}_{N-11},\mathcal{T}_{N1}$ and $\mathcal{T}_{N-12}$. The retarded Green's function elements 
are given by
\begin{align}
    G^{+}_{ll'}(\bar\mu)=\frac{(-1)^{l+l'}}{|\text{det}(\hat{Z})|} \big(\text{Cofact}[\hat{Z}]\big)_{l'l}~~,
\end{align}
where $\text{Cofact}[\hat{Z}]$ generates the cofactor of the matrix $\hat{Z}$. $\hat{Z}$ forms a tri-diagonal matrix in real space basis as follows:
\begin{align}
    Z_{ll'}=\begin{pmatrix}
        A&1&0&0&0&...&0&0&0\\
        1&2D&1&0&0&...&0&0&0\\
        0&1&\bar\mu&1&0&...&0&0&0\\
        0&0&1&\bar\mu&1&...&0&0&0\\
        ...&...&...&...&...&...&...&...&...\\
        0&0&0&0&0&...&\bar\mu&1&0\\
        0&0&0&0&0&...&1&2D&1\\
         0&0&0&0&0&...&0&1&A
    \end{pmatrix}_{N \times N}.
\end{align}
The determinant of the matrix $\hat{Z}$ is given by 
\begin{align}
     &\text{det}(\hat{Z})=(2AD-1)\big\{A(2DY_{N-4}-Y_{N-5})-Y_{N-4}\big\}\nonumber\\
    &~~~~~~~~~~~~~-A\big\{A(2DY_{N-5}-Y_{N-6})-Y_{N-5}\big\},
\end{align}
where $A=\bar\mu/2+i\sqrt{1-\bar\mu^2/4}$, $2D=\bar\mu-\gamma^2\big(\bar\mu-i\sqrt{4-\bar\mu^2}\big)/2$, and  $Y_N=\sin(N+1)k_F/\sin{k_F}$. Here, $k_F$ denotes the Fermi momentum and is defined via  $k_F=\cos^{-1}{(\bar\mu/2)}$. After evaluating the co-factors, 
 we get the required Green's function elements as follows: 
\begin{align}
    &G^{+}_{21}(\bar\mu)=-\frac{1}{\text{det}(\hat{Z})}\big\{A(2DY_{N-4}-Y_{N-5})-Y_{N-4}\big\},\\
    &G^{+}_{N1}(\bar\mu)=\frac{(-1)^{N+1}}{\text{det}(\hat{Z})},~~~~~G^{+}_{N-11}(\bar\mu)=\frac{(-1)^N}{\text{det}(\hat{Z})}A\\
    &G^{+}_{N-12}(\bar\mu)=\frac{(-1)^{N+1}}{\text{det}(\hat{Z})}A^2.
\end{align}
The corresponding transmission coefficients are  
\begin{align}   &\mathcal{T}_{21}=\gamma^2\frac{(4-\bar\mu^2)}{|\text{det}(\hat{Z})|^2}\Xi,~~~~\mathcal{T}_{N-11}=\gamma^2\frac{(4-\bar\mu^2)}{|\text{det}(\hat{Z})|^2},\\
&\mathcal{T}_{N1}=\frac{(4-\bar\mu^2)}{|\text{det}(\hat{Z})|^2},~~~~\mathcal{T}_{N-12}=\gamma^4\frac{(4-\bar\mu^2)}{|\text{det}(\hat{Z})|^2},
\end{align}
 where $\Xi$ is an oscillating function of $N$, and  defined by the following relation $\Xi=|A(2D\sin{(N-3)k_F}-\sin{(N-4})k_F)-\sin{(N-3)k_F}|^2/\sin^2{k_F}$. Utilizing these expressions, we finally get $\Delta_{\mu,N}$, $\tilde{G}_{N}$ and $\Delta j$ as follows :
\begin{align}
    &\Delta_{\mu,N}=\frac{2\pi|\text{det}(\hat{Z})|^2 I_g/[e (4-\bar\mu^2)\gamma^2]+(\Xi-1)\delta\mu/2 }{\Xi+1+2\gamma^2},\nonumber\\
  &\tilde{G}_{N}=\frac{ e^2 (4-\bar\mu^2) }{|\text{det}(\hat{Z})|^2 }\frac{(\gamma^2 \Xi+\gamma^2+2)(\Xi+1+2\gamma^2)-\gamma^2(\Xi-1)^2}{4\pi(2\gamma^2+\Xi+1)},\nonumber\\
     &\Delta j=2\frac{\Xi-1}{\Xi+1+2\gamma^2}I_g. \label{S112}
\end{align}
Eq.~\ref{S112} gives the periodic variation of $\Delta j$ with the system size $N$, which is displayed in Fig.~\ref{Fig2}(c). We find that whenever $\sin{(N-3)k_F}=0$, we get $\Delta j=0$ (irrespective of the value of $\gamma$), i.e., the nonreciprocity in the electrical current vanishes. In this limit 
\begin{align}
    &\Delta_{\mu,N}\Big|_{\Xi=1}=\frac{\pi|\text{det}(\hat{Z})|^2 I_g }{e (4-\bar\mu^2)\gamma^2(1+\gamma^2)}.
\end{align}
Therefore, the \textcolor{black}{chemical} potentials of the S and D baths are independent of $\delta\mu$. This happens whenever
 the distance between the S and D baths, i.e., ($N-3$) matches with an integer multiple of $\lambda_F/2$,  where $\lambda_F=2\pi/\cos^{-1}({\bar{\mu}/2})$.
 
\begin{figure}[h!]
\centering
\includegraphics[width=0.47\textwidth]{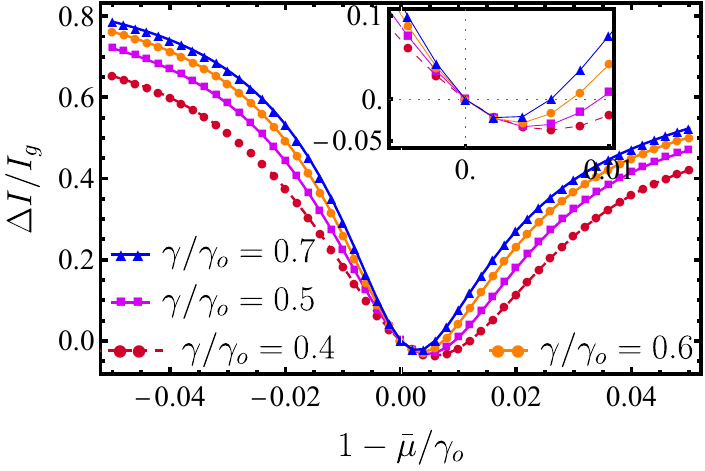}
\caption{ Control of current nonreciprocity using quantum coherence in a long wire with multiple S-D baths. The parameters are $N=60$, $\delta\mu/\gamma_o=0.1$, 
 and $I_g/I_o=0.1$. We fix $e=1$ in the plots.}\label{M60_SD_ballistic_nonreciprocity_fig}
\end{figure}

  Next, we investigate whether such coherence effects survive in a longer wire (with multiple dimers), where we have an extended ballistic region between consecutive S and D baths. In Fig.\ref{M60_SD_ballistic_nonreciprocity_fig}, we consider a wire of size $N=60$, where the S-baths are coupled at sites $x=2,8,..., N-4$ and the D-baths are coupled at sites $x=5,11,..., N-1$. We vary the average L-R \textcolor{black}{chemical} potential $\bar\mu$ while keeping $\delta\mu$ constant. We observe that the current nonreciprocity $\Delta I$ for different values of $\gamma$ becomes zero at $\bar\mu/\gamma_o=1.0$ (inset of the Fig.~\ref{M60_SD_ballistic_nonreciprocity_fig}).
\subsection{Change transport through an extended lattice with one S and D bath using master equation approach}\label{Appendix_One S-D in a long ballistic classical wire}
Here, we show that the oscillations and zeros of current nonreciprocity obtained within the quantum modeling do not appear in a classical transport channel within the master equation description.  We consider a lattice of size $N>4$.
The discrete time-evolution equations for the density fields $\rho^{\pm}(x,t)$ at the middle sites of the lattice, with only one S at $x=2a$ and one D at $x=(N-1)a$, are:
\begin{align}  
&\rho^{\pm}(2a,t+\tau)-\rho^{\pm}(2a,t)=\rho^{\pm}(2a\mp a,t)\nonumber\\
&~~~~~~~~~~~~~~~~~~~~~+(1-p)\rho^{\mp}(2a,t)-\rho^{\pm}(2a,t)+\frac{I_g \tau}{2ae},\\  
&\rho^{+}(x,t+\tau)-\rho^{+}(x,t)=\nonumber\\
&~~~~~~~~~~~~~~~~[1+(p-1)\delta_{x,3a}]\rho^{+}(x-a,t)-\rho^{+}(x,t),\\
&\rho^{-}(x,t+\tau)-\rho^{-}(x,t)=\nonumber\\
&~~~~~~~~~~~~[1+(p-1)\delta_{x,(N-2)a}]\rho^{-}(x+a,t)-\rho^{-}(x,t),\\
&\rho^{\pm}(L,t+\tau)-\rho^{\pm}(L,t)=\rho^{\pm}(L\mp a,t)\nonumber\\
&~~~~~~~~~~~~~~~~~~~~~~~~+(1-p)\rho^{\mp}(L,t)-\rho^{\pm}(L,t)-\frac{I_g \tau}{2ae},
 \end{align}
where, $x=3a,3a,...,(N-2)a$. 
The solutions in the steady state are obtained with the boundary condition $\rho^{+}(a)=\delta\rho+\rho_o$ and $\rho^{-}(L+a)=\rho_o$, and they are given by 
\begin{align}
   &\rho^{+}(2a)=\frac{(2-p)[\frac{I_g}{2ev_F}+\delta\rho]+(3-2p)\rho_o}{p(3-2p)},\\
      &\rho^{-}(2a)=\frac{\rho_o +2(1-p) [\frac{I_g}{2ev_F}+\delta\rho+\rho_o]}{p(3-2p)},\\
      &\rho^{+}(L)=\frac{(p-1)\frac{I_g}{ev_F}+\delta\rho+(3-2p)\rho_o}{p(3-2p)},\\
    &\rho^{-}(L)=\frac{(p-2)\frac{I_g}{2ev_F}+(1-p)\delta\rho+(3-2p)\rho_o}{p(3-2p)}.
\end{align}
The rest is obtained using: $\rho^{+}(L-a)=\rho^{+}(L-2a)=...=\rho^{+}(3a)=p\rho^{+}(2a)$, and $\rho^{-}(3a)=\rho^{-}(4a)=...=\rho^{-}(L-a)=p\rho^{-}(L)$. Therefore, the current entering (leaving) the lattice is
\begin{align}
    I_{\text{in}}=I_{\text{out}}=\frac{e v_F \delta\rho-(1-p) I_g }{3-2p},
\end{align}
and the current in the lattice between the S and D is 
\begin{align}   J(x)=\frac{e v_F \delta\rho+(2-p) I_g }{3-2p},~~~x=2a,3a,...,L-a.
\end{align}
We found that the current propagating through the lattice is independent of its length, and consequently, the nonreciprocity does not vary with $N$.

\section{Resistive circuit model of multiple S-D dimers}\label{resistive model}
We have started our introduction in Sec.~\ref{Sec_Intro} with a few S and D currents in three resistive circuits. 
Here, we focus on the resistive circuit model of the wire (length $L$) coupled to multiple S and D currents. Let us assume there are $N_i/2= N/2-1$ pairs of S and D currents arranged in the configuration (SDSD...) on the wire with a separation $ a$ between them. The resistance $R$ of the wire is uniformly distributed along its length, i.e.,  $R=L/\sigma$, where  $\sigma$ denotes the wire's conductivity. All the currents are taken due to the flow of positive charges ($e>0$). The current flowing into the wire for a forward voltage bias $V$ is
\begin{align}
   & I_{\text{in}}=\frac{\sigma V -( N/2-1) a I_g}{( N-1) a }=\sigma\frac{V }{L}-\frac{I_g}{2}\bigg(1-\frac{ a}{L}\bigg).\label{S125}
\end{align}
The current nonreciprocity becomes $I_g$ once again for $L\gg a$.
Next, we provide the voltage drop $V(x)$  at the locations $x$ in the wire where the S and D currents are coupled:
\begin{align}
      V(l{a}) 
    &=V-\frac{{a}}{L}\Bigg(V+\frac{I_g}{2}\frac{{a}}{\sigma}\Bigg)(l-2)-\frac{{a}}{L}\Bigg(V-\frac{I_g}{4}\frac{L-2{a}}{\sigma}\Bigg)\nonumber\\
    &~~~+(-1)^l \frac{I_g}{4}\bigg(\frac{{a}}{\sigma}\bigg),~~\text{for}~~l=2,3,... N-1.
\end{align}
The voltage bias across the boundaries give $V({a})=V$, and $V(L+{a})=0$. Now, the heat evolved per unit time
from the wire (Joule heating) due to the flow of charges between sites  $x=l{a}$ and $x=(l+1){a}$ is :
\begin{align}
     h_{l,l+1}= \frac{{a}}{\sigma}\bigg[\frac{\sigma V +{a} I_g/2}{L}+(-1)^{l}\frac{I_g}{2}\bigg]^2.
\end{align}
Therefore, the average heat energy generated (per unit time) from the bulk of the wire, over a length scale ${a}$, is :
\begin{align}
    \frac{ h_{l-1,l}+h_{l,l+1}}{2}= \frac{{a}}{\sigma}\bigg[\bigg(\frac{\sigma V +{a} I_g/2}{L}\bigg)^2+\bigg(\frac{I_g}{2}\bigg)^2\bigg].
\end{align}
The above expression looks somehow similar to the form in Eq.~\ref{heat_current_multiple dimers}, which was derived using S-D baths in the microscopic treatment.

\bibliography{references}

\end{document}